\pdfoutput=1

\documentclass[11pt,twoside,a4paper,cmspaper,final,collab]{cms-tdr}
\def\svnVersion{269923}\def\cmsCernNo{2013-037}\def\cmsCernDate{\today}\def\cmsMessage{Published in the European Physical Journal C as \href{http://dx.doi.org/10.1140/epjc/s10052-014-3149-z}{\doi{10.1140/epjc/s10052-014-3149-z.}}}

\begin{document}\cmsNoteHeader{EXO-13-008}

\hyphenation{had-ron-i-za-tion}
\hyphenation{cal-or-i-me-ter}
\hyphenation{de-vices}

\RCS$Revision: 262634 $
\RCS$HeadURL: svn+ssh://svn.cern.ch/reps/tdr2/papers/EXO-13-008/trunk/EXO-13-008.tex $
\RCS$Id: EXO-13-008.tex 262634 2014-10-01 15:21:19Z alverson $
\newlength\cmsFigWidth
\ifthenelse{\boolean{cms@external}}{\setlength\cmsFigWidth{0.85\columnwidth}}{\setlength\cmsFigWidth{0.4\textwidth}}
\ifthenelse{\boolean{cms@external}}{\providecommand{\cmsLeft}{top\xspace}}{\providecommand{\cmsLeft}{left\xspace}}
\ifthenelse{\boolean{cms@external}}{\providecommand{\cmsRight}{bottom\xspace}}{\providecommand{\cmsRight}{right\xspace}}
\ifthenelse{\boolean{cms@external}}{\providecommand{\suppMaterial}{the supplemental material [URL will be inserted by publisher]}}{\providecommand{\suppMaterial}{Appendix~\ref{app:suppMat}}}
\cmsNoteHeader{EXO-13-008}
\newcommand{\WR}{\ensuremath{\PW_{\cmsSymbolFace{R}}}\xspace}
\newcommand{\WL}{\ensuremath{\PW_{\cmsSymbolFace{L}}}\xspace}
\newcommand{\ZR}{\ensuremath{\cPZ_{\cmsSymbolFace{R}}}\xspace}
\newcommand{\ZL}{\ensuremath{\cPZ_{\cmsSymbolFace{L}}}\xspace}
\newcommand{\Nmu}{\ensuremath{\cmsSymbolFace{N}_{\Pgm}}\xspace}
\newcommand{\Nell}{\ensuremath{\cmsSymbolFace{N}_{\ell}}\xspace}
\newcommand{\Ne}{\ensuremath{\cmsSymbolFace{N}_{\Pe}}\xspace}
\newcommand{\Ntau}{\ensuremath{\cmsSymbolFace{N}_{\tau}}\xspace}
\providecommand{\re}{\ensuremath{\cmsSymbolFace{e}}}
\newcommand{\KL}{\ensuremath{\PK_{\cmsSymbolFace{L}}}\xspace}
\newcommand{\KS}{\ensuremath{\PK_{\cmsSymbolFace{S}}}\xspace}
\newcommand{\Mee}{\ensuremath{M_{\Pe \Pe}}\xspace}
\newcommand{\Mmumu}{\ensuremath{M_{\Pgm \Pgm}}\xspace}
\newcommand{\Mll}{\ensuremath{M_{\ell \ell}}\xspace}
\newcommand{\Mlljj}{\ensuremath{M_{\ell \ell j j}}\xspace}
\newcommand{\Meejj}{\ensuremath{M_{\Pe \Pe j j}}\xspace}
\newcommand{\Mmumujj}{\ensuremath{M_{\Pgm \Pgm j j}}\xspace}
\newcommand{\Memujj}{\ensuremath{M_{\Pe \Pgm j j}}\xspace}
\newcommand{\MWR}{\ensuremath{M_{\PW_{\cmsSymbolFace{R}}}}\xspace}
\newcommand{\Zpeak}{\ensuremath{\cPZ_\text{peak}}\xspace}
\providecommand{\FEWZ} {{\textsc{fewz}}\xspace}
\authorrunning{CMS}
\titlerunning{Search for heavy right-handed neutrinos and W bosons}
\title{Search for heavy neutrinos and $\PW$ bosons with right-handed couplings in proton-proton collisions at $\sqrt{s} = 8\TeV$}

\date{\today}

\abstract{
A search for heavy, right-handed neutrinos, \Nell ($\ell = \Pe, \mu$), and right-handed \WR\ bosons,
which arise in the left-right symmetric extensions of the standard model, has been performed by the CMS
experiment.  The search was based on a sample of two lepton plus two jet events collected in 
proton-proton collisions at a center-of-mass energy of 8\TeV corresponding to an integrated luminosity of 19.7\fbinv.
For models with strict left-right symmetry, and assuming only one \Nell flavor contributes significantly to the \WR decay width, the
region in the two-dimensional $(M_{\WR}, M_{\Nell})$ mass plane excluded at a 95\% confidence level
extends to approximately $M_{\WR} = 3.0\TeV$ and covers a large range of neutrino masses
below the \WR boson mass, depending on the value of $M_{\WR}$.  This search significantly extends
the $(M_{\WR}, M_{\Nell})$ exclusion region beyond previous results.}

\hypersetup{%
pdfauthor={CMS Collaboration},%
pdftitle={Search for heavy neutrinos and W bosons with right-handed couplings in proton-proton collisions at sqrt(s) = 8 TeV},%
pdfsubject={CMS},%
pdfkeywords={CMS, physics}}

\maketitle 

\section{Introduction}
\label{sec:intro}

The standard model (SM)~\cite{SM1,SM2,SM3} explicitly incorporates the parity violation observed in weak
interactions through the use of a left-handed chiral $SU_{L}(2)$ gauge group which includes
the left-handed gauge bosons $\WL^{\pm}$ and $\ZL$.  One of the attractive
features of left-right (LR) symmetric extensions~\cite{lr,lr1,lr2,lr3} to the standard model is that these models explain
parity violation in the SM as the consequence of spontaneous symmetry breaking of a larger
gauge group to $SU_{L}(2) \times SU_{R}(2)$ at a
multi-TeV mass scale.  The LR extensions introduce an additional right-handed $SU_{R}(2)$ symmetry
group to the SM, which includes heavy charged ($\WR^{\pm}$) and neutral (\ZR) gauge bosons
that could be produced at LHC energies.

In addition to addressing parity non-conservation in weak interactions, LR theories also provide
an explanation for the mass of SM neutrinos.  The observation of
neutrino oscillations~\cite{nu1,nu2} requires that neutrinos have mass, and the fact
that the neutrino mass scale~\cite{numass} is far below that of
quarks and charged leptons suggests that the origin of neutrino mass may be distinct from
the origin of mass for the other SM fermions.  Heavy right-handed Majorana neutrinos
(\Ne, \Nmu, and \Ntau), which are naturally present in LR models, provide a possible
explanation for the mass of SM neutrinos through the see-saw mechanism~\cite{seesaw1, seesaw2}.

We search for \WR bosons produced in a sample of proton-proton collisions at a
center-of-mass energy $\sqrt{s} = 8$\TeV and collected by the CMS detector at the CERN LHC.
This search, which expands upon a previous search
using $\sqrt{s} = 7$\TeV data~\cite{EXO-11-091}, assumes the production
of a \WR boson that decays to a
charged lepton (we consider $\ell = \Pe, \mu$) and to a right-handed neutrino \Nell.
The decay of the right-handed neutrino produces a second charged lepton of the same
flavor together with a virtual right-handed charged boson $\WR^*$.
When the $\WR^*$ decays to a pair of quarks, we arrive at the decay chain:

\begin{equation*}
\WR \to \ell_1 \Nell \to \ell_1 \ell_2 \WR^* \to \ell_1 \ell_2 \cPq \cPaq.
\end{equation*}

The quarks hadronize into jets ($j$), resulting in an observable final state containing two
same-flavor charged leptons and two jets.  Although the potential Majorana nature of the
right-handed neutrinos implies the final-state charged leptons can have the same sign, we
do not impose any charge requirements on the final-state electrons or muons in this analysis.

This search is characterized by the masses of the \WR boson (\MWR) and the right-handed
neutrino \Nell ($M_{\Nell}$), which are allowed to vary independently.
Although $M_{\Nell} > M_{\WR}$ is allowed in the
LR symmetric model, it is not considered in this analysis in favor of the dominant
$\cPq\cPaq' \to \WR$ production mechanism.
As the branching fraction for $\WR \to \ell \Nell$ depends on the
number of heavy-neutrino flavors accessible at LHC energies, results are first
interpreted assuming that only one neutrino flavor, namely \Ne or \Nmu, is
light enough to contribute significantly to the \WR boson decay width.
Results are then interpreted assuming degenerate \Ne, \Nmu, and \Ntau masses.

For given \WR boson and \Nell mass assumptions, the signal cross section can be predicted from the
value of the coupling constant $g_R$, which denotes the strength of the gauge interactions
of \WR bosons.  We assume strict LR symmetry, such that $g_R$ is equal to
the (left-handed) weak interaction coupling strength $g_L$ at $M_{\WR}$, and we also assume
identical quark and neutrino mixing matrices for the left- and right-handed interactions.
The \WR boson production cross section can then be calculated by
the \FEWZ program~\cite{kfactor} using the left-handed $\PWpr$ model~\cite{wpr}.
Finally, the left-right boson and lepton mixing angles are assumed to be small~\cite{mixAngles}.

The theoretical lower limit on \WR\ mass of $M_{\WR} \gtrsim 2.5$\TeV~\cite{mix1,mix2}
is estimated from the measured size of the $\KL \text{--} \KS$ mass difference.  Searches for $\WR \to \cPqt\cPaqb$ decays at
the LHC using $\sqrt{s} = 7$ and 8\TeV data~\cite{atlastb,cmstb,cmstb2} have excluded \WR boson masses below 2.05\TeV at 95\% confidence level (CL), and
previous searches for $\WR \to \ell \Nell$ at the LHC
excluded at 95\% CL a region in the two-dimensional parameter space $(M_{\WR},M_{\Nell})$
extending to nearly $M_{\WR} = 2.5$\TeV~\cite{atlas, EXO-11-091}.  This paper describes the first
direct search that is sensitive to \MWR values beyond the theoretical lower mass limit.

\section{The CMS detector}

The central feature of the CMS apparatus is a
superconducting solenoid, of 6\unit{m} internal diameter, providing a field
of 3.8\unit{T}. Within the field volume are the silicon pixel and strip
tracker, the PbWO$_4$ crystal electromagnetic calorimeter (ECAL) and the
brass and scintillator hadron calorimeter (HCAL). Muons are measured in
gas-ionization detectors embedded in the steel flux-return yoke.
The ECAL has an energy resolution of better than 0.5\%
for unconverted photons with transverse energies $\ET \equiv E / \cosh \eta > 100$\GeV.
The muons are measured in the
pseudorapidity window $|\eta|< 2.4$, where $\eta = -\ln [\tan(\theta/2)]$ and
$\theta$ is the polar angle with respect to the counterclockwise-beam direction.
The muon system detection planes are made of
three technologies: drift tubes, cathode strip chambers, and resistive-plate chambers.
Matching the muons to the tracks measured in the
silicon tracker results in a transverse momentum ($\pt \equiv | p | / \cosh \eta$)
resolution between 1 and 10\% for $\pt < 1$\TeV.  The inner tracker measures charged
particles within the range $|\eta| < 2.5$
and provides an impact parameter
resolution of $\sim$15~$\mu$m and a \pt
resolution of about 1.5\% for 100\GeV particles.  The first
level of the CMS trigger system, composed of custom hardware
processors, uses information from the calorimeters and muon detectors
to select up to 100\unit{kHz} of events of interest.
The high-level trigger (HLT) processor farm uses information from all
CMS subdetectors to further decrease the
event rate to about 400\unit{Hz} before data storage.
A more detailed description of the CMS detector, together with a
definition of the coordinate system used and the relevant
kinematic variables, can be found elsewhere~\cite{CMS}.

The particle-flow event reconstruction technique~\cite{pf1,pf2}
used to reconstruct jets in this analysis consists in reconstructing and
identifying each single particle with an optimized combination of all
subdetector information. The energy of photons is directly obtained from
the ECAL measurement, corrected for zero-suppression effects. The energy
of electrons is determined from a combination of the track momentum at
the main interaction vertex, the corresponding ECAL cluster energy, and
the energy sum of all bremsstrahlung photons attached to the track.
The energy of muons is obtained from the corresponding track momentum.
The energy of charged hadrons is determined from a combination of the
track momentum and the corresponding ECAL and HCAL energy, corrected
for zero-suppression effects and for the response function of the
calorimeters to hadronic showers. Finally, the energy of neutral hadrons
is obtained from the corresponding corrected ECAL and HCAL energy.

\section{Data and Monte Carlo samples}
\label{sec:datasets}

The search for \WR boson production described in this paper is performed using pp collision data
collected with the CMS detector at $\sqrt{s} = 8$\TeV in 2012.  The data sample
corresponds to an integrated luminosity of 19.7~\fbinv.
Candidate $\WR \to \Pe \Ne$ events are collected using
a double-electron trigger that requires two clusters in ECAL with $\ET > 33$\GeV each.
These ECAL clusters are loosely matched at the HLT stage to tracks formed from hits in
the pixel detector.  To reject hadronic backgrounds, only a small amount of energy in the
HCAL may be associated with the HLT electron candidates.
Muon channel events are selected with a single-muon trigger that requires at least
one candidate muon with $\pt > 40$\GeV and $\abs{\eta}< 2.1$, as reconstructed by the HLT.

Simulated $\WR \to \ell \Nell$ signal samples are generated assuming
$M_{\Nell} = \frac{1}{2} \MWR$ using \PYTHIA 6.4.26~\cite{pythia}, a tree-level
Monte Carlo (MC) generator, with CTEQ6L1 parton
distribution functions (PDF)~\cite{parton} and underlying event tune Z2*~\cite{tune}.
The MC generator includes the LR symmetric
model with the assumptions previously mentioned.  The final state leptons and jets in these
signal events are sufficiently energetic to allow reconstruction effects to be addressed apart
from the kinematic requirements discussed below.  With this separation,
the extension from $M_{\Nell} = \frac{1}{2} \MWR$ to the full two-dimensional
$(M_{\WR},M_{\Nell})$ mass plane for signal events is straight-forward,
as is discussed in Section~\ref{sec:limits}.
The dominant backgrounds to \WR boson production include SM processes with at least two charged
leptons with large transverse momentum, namely $\ttbar \to \cPqb \PWp \cPaqb \PWm$ and
Drell--Yan (DY)+jets processes.  All remaining SM background events, which collectively
contribute less than 10\% to the total background level, are dominated by diboson and
single top quark processes.  The \ttbar background is estimated using control
samples in data and a simulated sample of fully leptonic \ttbar decays, which are
generated using the tree-level matrix element
MC generator \MADGRAPH 5.1.4.8~\cite{madgraph}.  The DY+jets background is estimated using
exclusive DY+n jets ($n=0$, 1, 2, 3, 4) simulated samples generated with \MADGRAPH 5.1.3.30.
For the above \MADGRAPH samples, parton showering, fragmentation, and the underlying event
are handled by \PYTHIA.
A statistically comparable sample of DY+jets events generated with the tree-level MC
event generator \SHERPA 1.4.2~\cite{sherpa}, which incorporates parton showering and other effects
in addition to the hard process, is used to help quantify the systematic uncertainty in
the DY+jets background estimation.  Simulated diboson ($\PW \PW$, $\PW$Z, and ZZ) events
are generated using \PYTHIA 6.4.26, with the additional small contributions from diboson
scattering processes generated with \MADGRAPH 5.1.3.30.
The simulated single top quark (namely, tW) background sample is generated
via the next-to-leading-order MC generator \POWHEG 1.0~\cite{powheg1,powheg2,powheg3,powheg4}.
Parton showering and other effects are handled by \PYTHIA for the diboson and single top quark
background samples.

The generated signal and SM background events pass through a full CMS detector
simulation based on \GEANTfour \cite{geant4}, and are reconstructed
with the same software used to reconstruct collision data, unless otherwise noted.  The simulation 
is compared to data using various control samples, and when necessary the 
simulation is adjusted to account for slight deviations seen with respect to data.
Additional pp collisions in the same beam crossing (pileup) are also included for each
simulated event to realistically describe the $\sqrt{s} = 8$\TeV collision environment.

\section{Event selection and object reconstruction}
\label{sec:selection}

We assemble \WR boson candidates from the two highest-\pt (leading) jets
and two highest-\pt same-flavor
leptons (electrons or muons) reconstructed in collision data or simulation events.
Candidate events are first selected by the CMS trigger system using the lepton triggers
described previously.  The electron and muon trigger efficiencies are determined using the
``tag and probe'' techniques applied to $\cPZ \to \ell \ell$ candidates~\cite{zprime7TeV,muid,eid}.  Simple triggers, requiring a single ECAL cluster with $\ET > 300$\GeV, collected events
with high-\pt electrons to help evaluate the trigger efficiency for electron channel events with
high dielectron mass~\cite{zprime8TeV}.  Following the application of object and event selection
requirements mentioned below, the trigger efficiency for $\WR \to \ell \Nell$ candidate
events is greater than 99\% (98\%) in the electron (muon) channel.

Because of the large expected mass of the \WR boson, electron and muon reconstruction and
identification are performed using algorithms optimized for objects with large
transverse momentum~\cite{zprime7TeV,zprime8TeV}.  Non-isolated muon
backgrounds are suppressed by computing the transverse momentum sum of all additional
tracks within a cone of $\Delta R < 0.3$ about the muon direction,
where $\Delta R = \sqrt{\smash[b]{(\Delta \eta)^{2} + (\Delta \phi)^{2}}}$
(azimuthal angle $\phi$ in radians), and requiring the \pt sum to be less than
10\% of the muon transverse momentum.  This isolation requirement is only weakly
dependent on the number of pileup collisions in the event, as tracks with a large $\Delta{z}$ separation
from the muon, \ie, tracks from other $\Pp\Pp$ collisions, are not included in the isolation sum.
Electrons are expected to have minimal associated HCAL energy and also to appear isolated in both
calorimeters and in the tracker.  To minimize the effects of pileup,
electrons must be associated with the primary vertex,
which is the collision vertex with the highest $\sum \pt^{2}$ of all associated tracks.
As pileup collisions also produce extra energy in the calorimeters and can
make the electron appear non-isolated, calorimeter isolation for electron
candidates is corrected for the average energy density in the event~\cite{pileup2}.

Jets are reconstructed using the anti-\kt clustering
algorithm~\cite{antikt} with a distance parameter of 0.5.
Charged and neutral hadrons, photons, and leptons reconstructed with the CMS particle-flow
technique are used as input to the jet clustering algorithm.
To reduce the contribution to jet energy from pileup collisions, charged hadrons that do not
originate from the primary vertex in the event are not used in jet clustering.
After jet clustering, the pileup calorimeter energy contribution from neutral
particles is removed by applying a residual average area-based correction~\cite{pileup1,pileup2}.
Jet identification requirements~\cite{jetid} suppress jets from calorimeter noise and beam
halo, and the event is rejected if either of the two highest-\pt jet
candidates fails the identification criteria.
The jet four-momenta are corrected for zero-suppression effects and
for the response function of the calorimeters to hadronic showers
based on studies with simulation and data~\cite{jetpaper}.
As the electrons and muons from \WR boson decay are likely to be spatially separated from jets in
the detector, we reject any lepton found within a cone of radius
$\Delta R < 0.5$ from the jet axis for either of the two leading jets.

After selecting jets and isolated electrons or muons in the event,
$\WR \to \ell \Nell$ candidates are formed using the two leading
same-flavor leptons and the two leading jets that satisfy the
selection criteria.  The leading (subleading) lepton is required to have $\pt > 60\;(40)$\GeV, while
the \pt of each jet candidate must exceed 40\GeV.  Electrons and jets
are reconstructed within the tracker acceptance ($\abs{\eta}< 2.5$).  Muon acceptance extends to
$\abs{\eta}< 2.4$, although at least one muon is restricted to $\abs{\eta}< 2.1$ in order to be selected by the trigger.

We perform a shape-based analysis, searching for evidence of \WR boson production using
the four-object mass distribution (\Mlljj), where we consider events with
$\Mlljj > 600$\GeV.
To reduce the contribution from DY+jets and other SM backgrounds,
we also impose a requirement of $\Mll > 200$\GeV on the mass of the lepton pair
associated with the \WR boson candidate.

The decay of a \WR boson tends to produce final-state objects that have high \pt and
are separated in the detector.  We define the signal acceptance to include the
kinematic and detector acceptance requirements for the leptons and jets, lepton-jet separation,
and the minimum \Mll and \Mlljj requirements.  This signal acceptance, typically
near 80\% at $M_{\Nell} \sim M_{\WR}/2$, varies by less than 1\% between the electron and
muon channels because of differences in
detector acceptance for leptons.  Provided that the \WR boson decay
satisfies acceptance requirements, the ability to reconstruct
all four final-state particles is near 75\%\;2.8
(85\%) for the electron (muon) channel, 
with some dependence on \WR boson and \Nell masses.  However, if the mass of the
\WR boson is sufficiently heavy compared to that of the right-handed neutrino,
the $\Nell \to \ell jj$ decay products tend to overlap and it becomes
difficult to reconstruct two distinct jets or find leptons outside of the jet cone.
As a result, the signal acceptance as a function of $M_{\Nell}$ decreases
rapidly as $M_{\Nell}$ drops below about 10\% of the \WR boson mass.

\section{Standard model backgrounds}
\label{sec:bg}

The \ttbar background contribution to the $\Pe \Pe jj$ and $\mu \mu jj$ final states
is estimated using a control sample of $\Pe \mu jj$ events reconstructed in data.
Studies of simulated $\ttbar \to \Pe \Pe jj$, $\mu \mu jj$, and $\Pe \mu jj$ decays
confirm that the \Meejj and \Mmumujj distributions can be modeled by the \Memujj
distribution, so we apply selection requirements to $\Pe \mu jj$ events
that parallel those applied to electron and muon channel events.
The $\Pe \mu jj$ events are collected using the same HLT selection as $\mu \mu j j$ events,
although in this case only one muon is available for selection by the trigger.  This sample is
dominated by \ttbar events, and small contributions from other SM processes are
subtracted using simulation.  The relative fractions of
$\ttbar \to \Pe \Pe jj$, $\mu \mu jj$, and $\Pe \mu jj$
events that pass the selection criteria are determined from simulation.
Using this information, the \Memujj distribution for the $\Pe \mu jj$ control sample from
data is scaled to match the
expected \ttbar background contribution to the \Meejj and \Mmumujj distributions.
The scale factor derived from simulation is determined after requiring
$M_{\Pe \mu} > 200$\GeV and $\Memujj > 600$\GeV, which is equivalent to the
third and final selection stage in Table~\ref{tab:numbers_select}.  The scale factors
for the \ttbar background sample are $0.524 \pm 0.007$ and
$0.632 \pm 0.008$ in the electron and muon channels, respectively, where
the uncertainty in the values reflects the number of simulated \ttbar events
that satisfy all object and event requirements.  The trigger efficiency for
$\Pe \mu j j$ events is over 90\% for events with central muons
($\abs{\eta}< 0.9$) and decreases for events with more forward muons.  Consequently,
both the electron and muon scale factors are larger than the expected value of 0.5, given the
relative branching fractions for $\ttbar \to \Pe \Pe jj$, $\mu \mu jj$, and $\Pe \mu jj$
decays.

The \ttbar scale factors, determined from simulation, are checked using control regions in data.
We first consider events in both simulation and data where one or both jets
are identified as originating from a bottom quark.  After all selection requirements
are applied, reconstructed \ttbar decays dominate the event samples.  Accounting for contributions from
other SM processes using simulation, we compute scale factors
for $\Pe \mu j j$ events in data with $60 < M_{\Pe \mu} < 200$\GeV to estimate the \ttbar contribution
to the SM background when one or both jets are tagged as b jets using the medium working point
of the combined secondary vertex tagging algorithm~\cite{btag}.  The \Mee and \Mmumu distributions
in b-tagged data agree with expectations based on simulation and the $\Pe \mu j j$ control sample, and
the derived scale factors agree with those obtained from simulation within statistical precision.
For another cross-check, we compute the scale factor based on the expectation
that $\ttbar \to \Pe \mu j j$ should be twice the rate of $\ttbar \to \Pe \Pe jj$ or $\ttbar \to \mu \mu jj$.
Deviation from this expected ratio depends primarily on the differences in electron and muon
reconstruction and identification efficiencies.  The number of electron and muon channel
events in data in the $120 < \Mll < 200\GeV$ control region are thus used to derive the relative
efficiency difference between electrons and muons and then extract the
\ttbar scale factors.  The scale factors determined from this control region in data
are consistent with those derived from simulation, and the larger statistical
uncertainty (2\%) of this cross-check is taken as the systematic uncertainty
in the \ttbar normalization.

\begin{figure}[!htbp]
\centering
\includegraphics[width=0.45\textwidth]{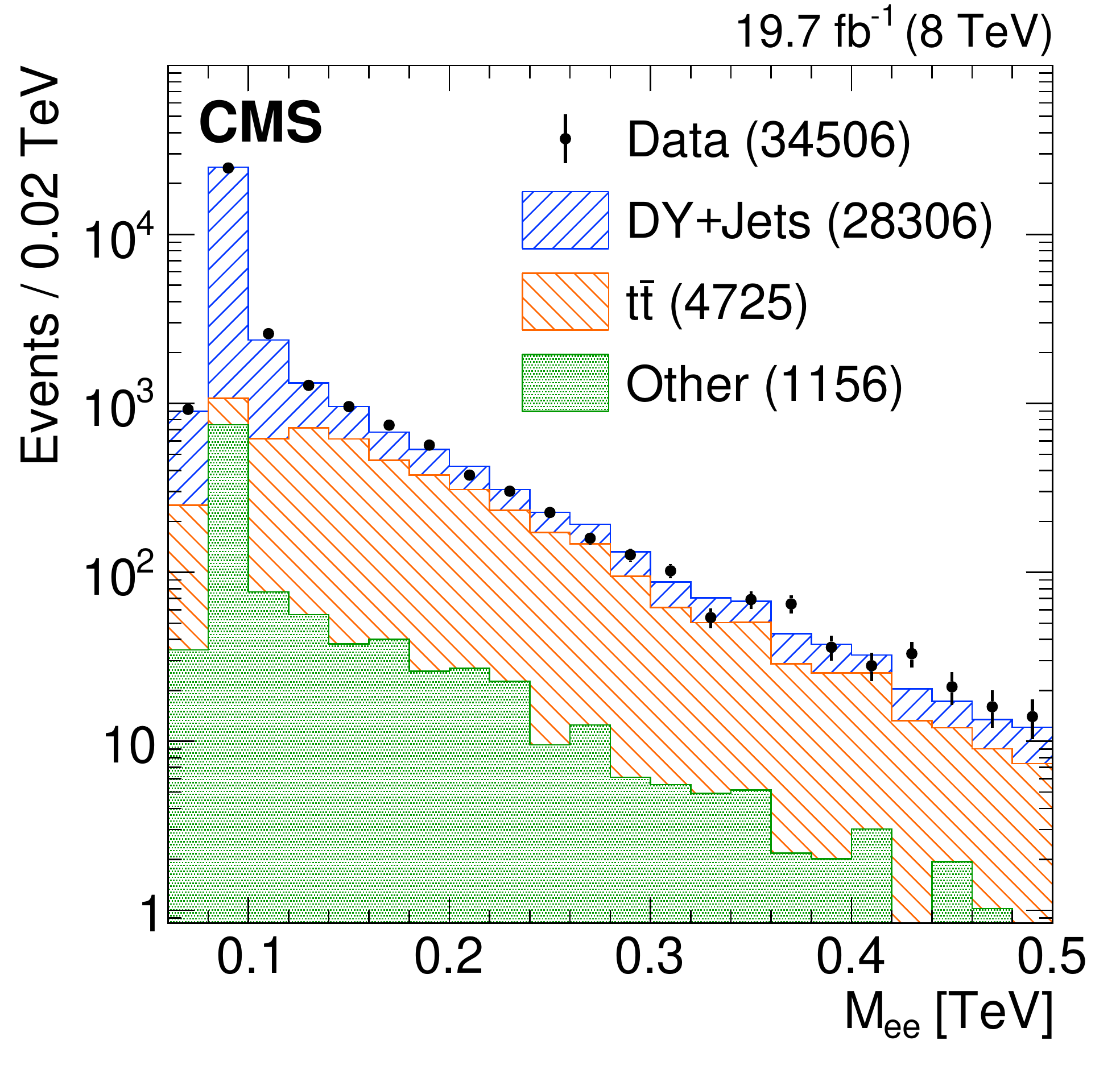}
\hfill
\includegraphics[width=0.45\textwidth]{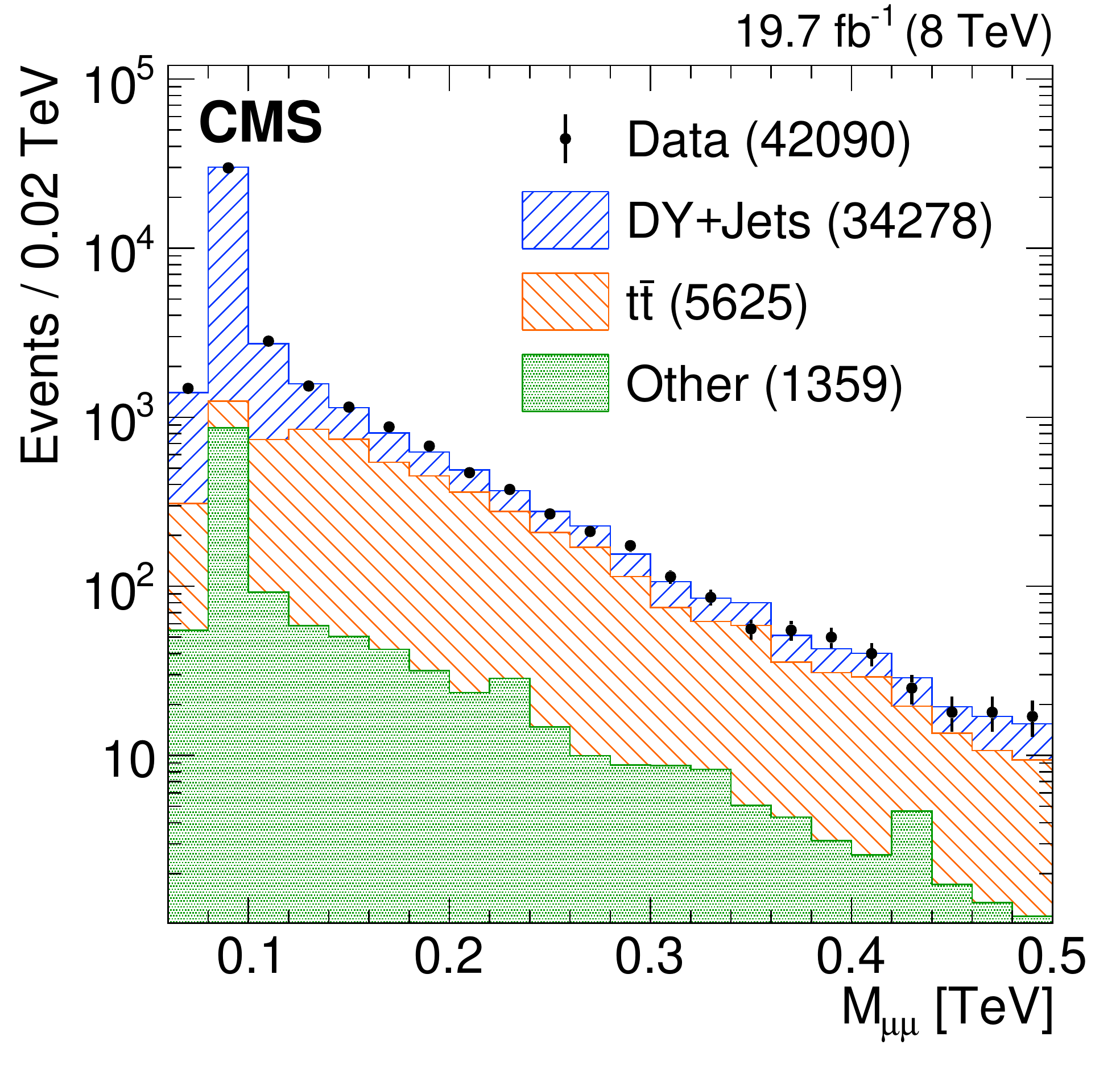}
\caption{Distribution of the invariant mass \Mee (\cmsLeft) and
\Mmumu (\cmsRight) for events in data (points with error bars)
with $\pt > 60\;(40)$\GeV for the leading (subleading) lepton
and at least two jets with $\pt > 40$\GeV, and for background contributions
(hatched stacked histograms) from data control samples (\ttbar) and simulation.
The numbers of events from each SM process are included in parentheses in the legend, where
the contributions from diboson and single top quark processes have been collected in the
``Other'' background category.
\label{fig:mLL}
}
\end{figure}

The DY+jets background contribution is estimated from
$\cPZ/\gamma^*\to \ell\ell$ decays reconstructed
in simulation and data.  The simulated DY+jets background
contribution is normalized to data using events in the dilepton mass region
$60 < \Mll < 120\GeV$ after kinematic requirements are applied on the
leptons and jets, which is the first selection stage indicated in
Table~\ref{tab:numbers_select}.  After removing the small contribution from
other SM background processes, the simulated DY+jets distributions are normalized to data
using scale factors of $1.000 \pm 0.007$ and $1.027 \pm 0.006$ for the
electron and muon channels, respectively, relative to inclusive
next-to-next-to-leading-order cross section calculations.  The uncertainty in this value
reflects the number of events from data with $60 < \Mll < 120\GeV$.
The shape of the \Mll distribution in data is in agreement
with SM expectations for $\Mll > 60$\GeV, as shown in Fig.~\ref{fig:mLL}.

The diboson and single top quark contributions to the total background are estimated
from simulation, based on
next-to-leading-order~\cite{MCFM} and approximate next-to-next-to-leading-order~\cite{Kidonakis}
production cross sections, respectively.  The background from \PW+jets processes, also estimated from
simulation, is negligible starting from the earliest selection stage.
Finally, the background contribution from multijet
processes is estimated using control samples in data and is also
found to be negligible at every selection stage.

The observed and expected numbers of events surviving
the selections are summarized in Table~\ref{tab:numbers_select},
which explicitly lists the contributions from \ttbar
and DY+jets processes while including all other SM background contributions in a single column.
The yields reflect the numbers of background events surviving each
selection stage, with normalization factors obtained from
simulation and control sample studies or taken directly from simulation.
The numbers of events observed at each selection stage agree with SM expectations
in both channels.

\begin{table*}[!htb]
\centering
  \topcaption{The total numbers of events reconstructed in data,
    and the expected contributions from signal and background samples,
    after successive stages of the selection requirements are applied.  For the
    first selection stage, all kinematic and identification requirements are
    imposed on the leptons and jets as described in the text.  The ``Signal''
    column indicates the expected contribution
    for $M_{\WR} = 2.5$\TeV, with $M_{\Nell} = 1.25$\TeV.  The ``Other''
    column represents the combined background contribution from diboson and single top quark
    processes.
    The uncertainties in the background
    expectation are derived for the final stage of selection and
    more details are given in Section~\ref{sec:results}.
    The total experimental uncertainty is
    summarized in the first signal uncertainty, and the second signal uncertainty
    represents the PDF cross section uncertainty.
    The yields from earlier stages of the selection
    have greater relative uncertainty than that for the
    final $\Mlljj > 600$\GeV selection stage.
}

\begin{tabular}{ccccccc} \hline
         &              &                &   \multicolumn{4}{c}{SM Backgrounds}\\
         \cline{4-7}
         &  Data        &     Signal     &   \multicolumn{1}{c}{Total}       &     \ttbar     &     DY+jets     &     Other    \\ \hline
Two electrons, two jets & 34506  &       30      &      34154     &       4725     &      28273     &      1156    \\
$\Mee > 200$\GeV       &  1717  &       29      &       1747     &       1164     &        475     &       108    \\
$\Meejj > 600$\GeV     &   817  & $ 29 \pm 1 \pm 3 $ & $ 783 \pm 51 $ & $ 476 \pm 42 $ & $ 252 \pm 24 $ & $ 55 \pm 12 $ \\ \hline
Two muons, two jets    & 42090  &       35      &      41204     &       5625     &      34220     &      1359    \\
$\Mmumu > 200$\GeV    &  2042  &       35      &       2064     &       1382     &        549     &       133    \\
$\Mmumujj > 600$\GeV  &   951  & $ 35 \pm 1 \pm 4 $ & $ 913 \pm 58 $ & $ 562 \pm 50 $ & $ 287 \pm 26 $ & $ 64 \pm 12 $ \\ \hline
\end{tabular}

\label{tab:numbers_select}
\end{table*}

\section{The \texorpdfstring{\MWR}{M[WR]} distribution and systematic uncertainties}
\label{sec:results}

Once all object and event selection criteria are applied, the \Mlljj distributions in
data and simulation are used to search for evidence of \WR boson production, where
the expected SM \Mlljj distribution is computed as the sum of the individual
background \Mlljj distributions.
The \Mlljj distribution is
measured in 200\GeV wide bins up to 1.8\TeV, as this bin width is
comparable to the mass resolution of the \WR boson for $M_{\WR} < 2.5$\TeV.
Beyond 1.8\TeV, events
are summed in two bins, $1.8 < \Mlljj < 2.2$\TeV and $\Mlljj > 2.2$\TeV, to
account for the small number of background events in the simulated and data control samples
at high mass.  The \Mlljj distributions for DY+jets,
diboson, and single top quark processes are taken from simulation, with the normalization
of each distribution as discussed previously.  The \Memujj distribution from data is
used to model the \ttbar background contribution in the electron and muon channels.

In our previous search for $\WR \to \mu \Nmu$ production using 7\TeV
collision data~\cite{EXO-11-091}, we
modeled the shape of each background \Mmumujj distribution using an exponential
lineshape.  For this search, we again find that an exponential function can be used to
describe each background \Mlljj distribution below 2\TeV, but
these \Mlljj distributions begin to deviate from the assumed exponential shape at high mass.
As a result, in this updated search
we use the \Mlljj distributions from each background process directly instead of relying
on exponential fits to model the shape of the SM backgrounds.

As the \ttbar background shape is taken from a control sample of $\Pe \mu j j$ events in data,
we examine the shape of the \ttbar background \Memujj distributions in both simulation and
data.  Based on the method to extract the background shape in our earlier search, we
fit each \Memujj distribution to an exponential lineshape for events surviving
all selection criteria for $\Pe \mu jj$ events.  The \ttbar background distribution
is again expected to decrease exponentially as \Mlljj increases, although we
allow for deviations at high mass (beyond 2\TeV) where the DY+jets background is more
significant.  The simulated \Memujj distribution
agrees with the exponential lineshape for $\Memujj < 2$\TeV, as expected, while we find that
the \Memujj distribution in the data control sample noticeably deviates from fit expectations
for $1.0 < \Memujj < 1.2\TeV$.  While the fit expects 94 events, only 78 events are
found in data in this region.  As a result, we correct the \Memujj distribution
from the data control sample to the
expected number of events from the exponential fit for $1.0 < \Memujj < 1.2\TeV$, and this
correction is reflected in Table~\ref{tab:numbers_select}.  The size of the correction
is taken as a systematic uncertainty in the shape of the \ttbar \Mlljj distribution.

\begin{figure}[!htbp]
\includegraphics[width=0.45\textwidth]{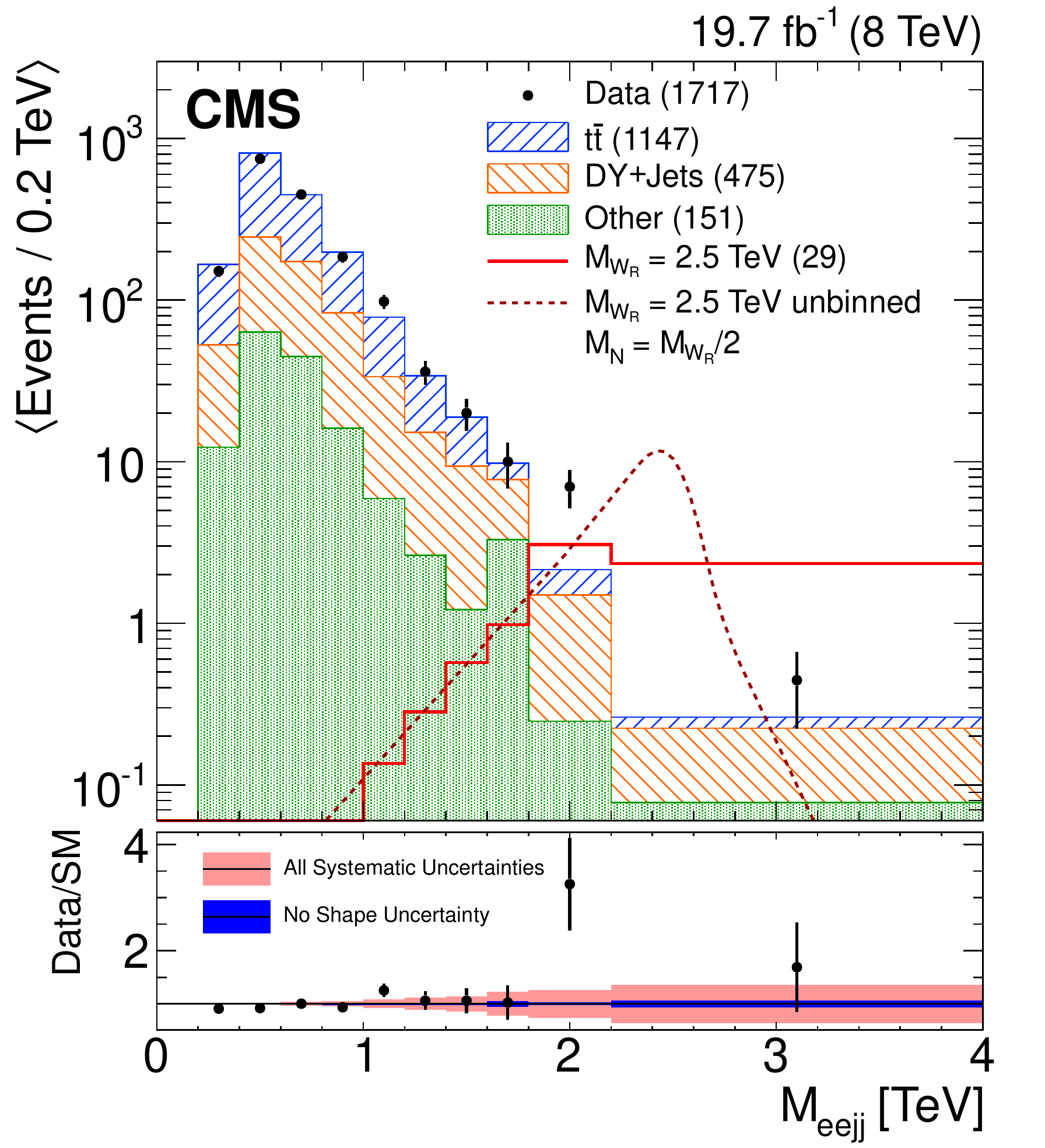}
\hfill
\includegraphics[width=0.45\textwidth]{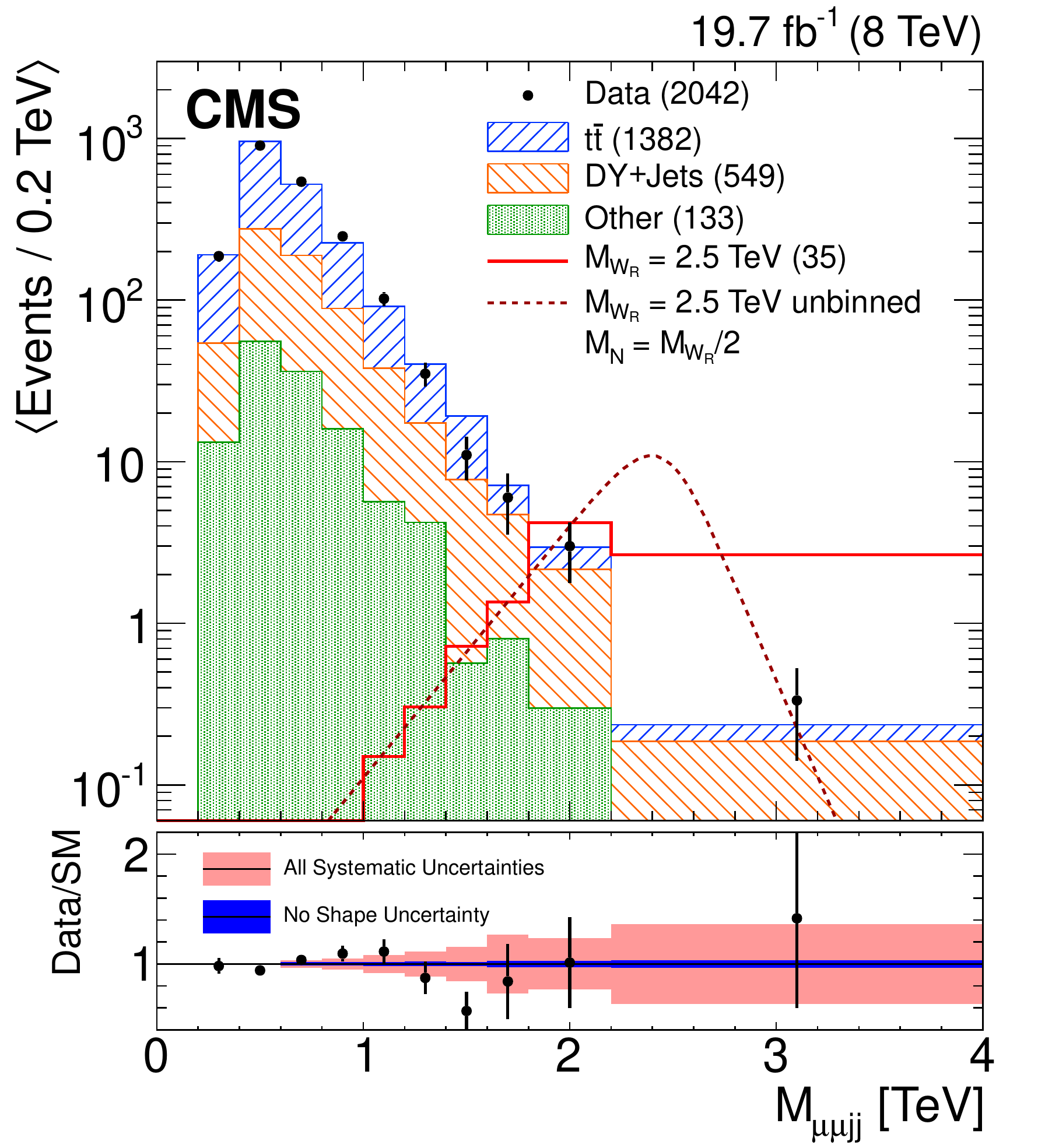}
\caption{Distribution of the invariant mass \Meejj (\cmsLeft) and
\Mmumujj (\cmsRight) for events in data (points with error bars)
with $\Mll > 200$\GeV and for background
contributions (hatched stacked histograms) from data
control samples (\ttbar) and simulation.
The signal mass point $M_{\WR}=2.5$\TeV, $M_{\Nell}=1.25$\TeV,
is included for comparison (open red histogram,
and also as a dotted line for the unbinned signal shape).
The numbers of events from each background process
(and the expected number of signal events) are
included in parentheses in the legend, where the contributions
from diboson and single top quark processes have been collected in the
``Other'' background category.
The data are compared with SM expectations in the lower portion
of the figure.  The total background uncertainty (light red band)
and the background uncertainty after neglecting
the uncertainty due to background modeling
(dark blue band) are included as a function of
\Mlljj for $\Mlljj > 600$\GeV (dashed line).
\label{fig:evtsPostMll}
}
\end{figure}

The \Mlljj distributions for events satisfying all selection criteria appear in Fig.~\ref{fig:evtsPostMll}.
A comparison of the observed data to SM expectations yields a
normalized $\chi^2$ of 1.4~(0.9) for electron (muon) channel events.
We observe an excess in the electron channel in the region $1.8
< \Meejj < 2.2\TeV$, where 14 events are observed compared to 4 events
expected from SM backgrounds.  This excess has a local significance of
2.8$\sigma$ estimated using the method discussed in
Section~\ref{sec:limits}.  This excess does not appear to be
consistent with $\WR \to \Pe \Ne$ decay.  We examined additional
distributions for events with $1.8 < \Meejj < 2.2\TeV$, including the
mass distributions $M_{\Pe jj}$ (for both the leading and subleading
electrons), $M_{\Pe\Pe}$, and $M_{jj}$, as well as the \pt
distributions for each of the final state particles.  In this
examination, we find no compelling evidence in favor of the signal
hypothesis over the assumption of an excess of SM background events in
this region.  Examining the charge of the electrons used to build \WR
boson candidates in data events with $1.8 < \Meejj < 2.2\TeV$, we find
same-sign electrons in one of the 14 reconstructed events.  In this
region, the same-sign SM background is expected to be on the order of
half an event due to SM diboson processes and charge misidentification
in DY+jets events.  No same-sign events are observed in the same mass
region of the distribution for the muon channel. For comparison,
making plausible assumptions for the properties of a signal
contributing in this region, one would expect half of the additional
events to have electrons with the same sign.

The uncertainties in modeling the shape of the background \Mlljj distributions
dominate the background systematic uncertainty, as shown in Fig.~\ref{fig:evtsPostMll}.  The background
\Mlljj uncertainty is determined in each mass bin based on the number of
events surviving all selection criteria for each background sample.  For the two
dominant backgrounds, an additional shape uncertainty is included as part of the background
shape uncertainty.

The additional \ttbar shape uncertainty is included for the $1.0 < \Mlljj < 1.2$\TeV mass region
based on the previously discussed correction to the \Memujj distribution for $1.0 < \Memujj < 1.2$\TeV.
No additional \ttbar shape uncertainty is applied at other \Mlljj values as the \Memujj distributions
in both data and simulation agree with the assumed exponential lineshape below 1.8\TeV, and the
statistical uncertainty of the $\Pe \mu j j$ control sample dominates at high mass.
For the DY+jets background, the \Mlljj shape uncertainty is determined using simulated 
samples from two different MC generators, \MADGRAPH and \SHERPA.
The difference between these two \Mlljj distributions, computed as a function of mass,
is taken as an additional systematic uncertainty in the DY+jets shape.

The uncertainty associated with the background normalization is taken
as the quadratic sum of the uncertainty in the scale factors determined from the cross-check
for \ttbar background performed on a control region in data, the uncertainty estimated from the
difference in the values obtained for DY+jets scale factors in the electron and muon channels, and the combined
cross section and luminosity uncertainties for the remaining backgrounds.
This overall background normalization uncertainty is small compared to the uncertainties
determined for the background shape.

Lepton reconstruction and identification uncertainties, which also contribute to
the total signal and background systematic uncertainty, are determined using
$\cPZ \to \Pe\Pe, \mu\mu$ events
reconstructed in both data and simulation.  Uncertainties in the jet and lepton energy
scales and resolutions also contribute to the systematic uncertainty.  These
uncertainties dominate the signal efficiency uncertainty, resulting in a
total systematic uncertainty of up to 10\% for the signal efficiency,
depending on the \WR boson mass assumption.  The combination of lepton and jet
energy scale, resolution, and efficiency uncertainties is less than
5\% for the background estimates taken from simulation.

The systematic uncertainties related to pileup, uncertainties in the proton PDFs, and initial- or final-state radiation are computed for the simulated background samples and are found to be small when compared to the 
background shape uncertainty.  Additional theoretical uncertainties for the SM background processes are covered by 
the shape uncertainty.  The total uncertainty for signal and background is determined for the final selection stage and
presented in Table~\ref{tab:numbers_select}.  Figure~\ref{fig:evtsPostMll}
summarizes the background uncertainty as a function of \Mlljj and displays the dominant background shape
uncertainty relative to the total background uncertainty.

\section{Limits on \texorpdfstring{\WR}{W[R]} boson production}
\label{sec:limits}

We estimate limits on \WR boson production using a multibin $\mathrm{CL}_\mathrm{S}$ limit
setting technique~\cite{cls1,cls2,Moneta:2010pm}.  The \Mlljj distributions obtained
from signal MC, each of the SM backgrounds, and the observed data all serve as limit inputs.
The systematic uncertainties mentioned previously are included
as nuisance parameters in the limit calculations.  We estimate the 95\% CL upper limit on the
\WR boson cross section multiplied by the $\WR \to \ell \ell jj$ branching fraction as
a function of \MWR and $M_{\Nell}$.
These results (available in tabular form in \suppMaterial) can be used for the
evaluation of models other than those considered in this paper.

The limits are computed for a set of \WR boson and \Nell mass assumptions, where $M_{\WR}$ starts at
1\TeV and increases in 100\GeV steps and the \Nell mass is taken
to be half the \WR boson mass.  For these determinations,
the \WR boson signal samples include the full CMS detector simulation.

The procedure to determine the limits on \WR boson production for a range of \Nell
mass assumptions ($M_{\Nell} < M_{\WR}$) proceeds as follows.  For a fixed value of \MWR,
the limits on $\WR \to \ell \Nell \to \ell \ell j j$
are determined as a function of $M_{\Nell}$ (up to \MWR) based on differences in kinematic acceptance, 
lepton-jet overlap, and \Mlljj shape relative to $M_{\Nell} = \frac{1}{2} M_{\WR}$.  As mentioned previously,
the combined reconstruction and identification efficiency for the \WR boson
and \Nell decay products varies by $\mathcal{O}(1\%)$ as a function of \MWR
once acceptance requirements are satisfied.  Consequently, for $M_{\Nell}$ values other than $M_{\Nell} = \frac{1}{2} M_{\WR}$, 
the \WR boson production cross section limits are computed using information from signal samples that do not include the
simulated detector response.

The cross section limit calculation based on the kinematic acceptance is compared with the
results for fully simulated samples using a spectrum of \Nell mass assumptions
for $\MWR =1$, 1.5, 2, and 3\TeV.
The difference between the two methods is at the percent level or smaller
for $M_{\Nell}$ masses greater than
10--20\% of the generated \WR boson mass.  Differences grow to $\mathcal{O}(10)$\% for
lighter right-handed neutrinos.
The ratio of the products of efficiency and acceptance for the two approaches
is computed as a function of
$M_{\Nell} / \MWR$, and a global fit to this distribution is used to
correct the cross section limits determined
as a function of $M_{\Nell}$ for all \MWR values.

The uncertainty in this correction is computed using the maximum difference in the efficiency times acceptance
ratio for the set of simulated samples as a function of $M_{\Nell} / \MWR$, unless the statistical uncertainty
in the ratio calculation dominates.  The impact of this uncertainty on
signal acceptance is propagated to the cross section limit calculations.  The overall effect on the limits
from this uncertainty is negligible for most $M_{\Nell}$ values, but can degrade the cross section limit by
5--10\% for \Nell masses below 10\% of \MWR.

Finally, we account for variations in the shape of the \Mlljj distribution.
As $M_{\Nell} \to 0$, neutrino production via a virtual \WR boson becomes
more significant.  As a result, the shape of the signal \Mlljj distribution is expected to
vary as a function of both \MWR and $M_{\Nell}$.  This effect is included in the limit calculations.

The largest uncertainty related to the $\WR \to \ell \Nell$ production estimation arises
from the variation in the predicted signal production cross section
as a result of the uncertainties in the proton PDFs, where we use the CTEQ6L1 PDF set for signal events.
The cross section uncertainty, which is not considered in the limit calculations,
ranges from 5\% for $M_{\WR} = 1$\TeV to 26\% for $M_{\WR} = 3$\TeV and is computed
following the PDF4LHC prescriptions~\cite{pdf4lhc,pdf4lhc2} for the CT10~\cite{ct10},
MSTW2008~\cite{mstw}, and NNPDF2.1~\cite{nnpdf} PDF sets.
The PDF uncertainties in the signal acceptance, which are small compared to the systematic
uncertainties for signal events mentioned previously, are included in the limit calculations.

\begin{figure}[!htbp]
\includegraphics[width=0.45\textwidth]{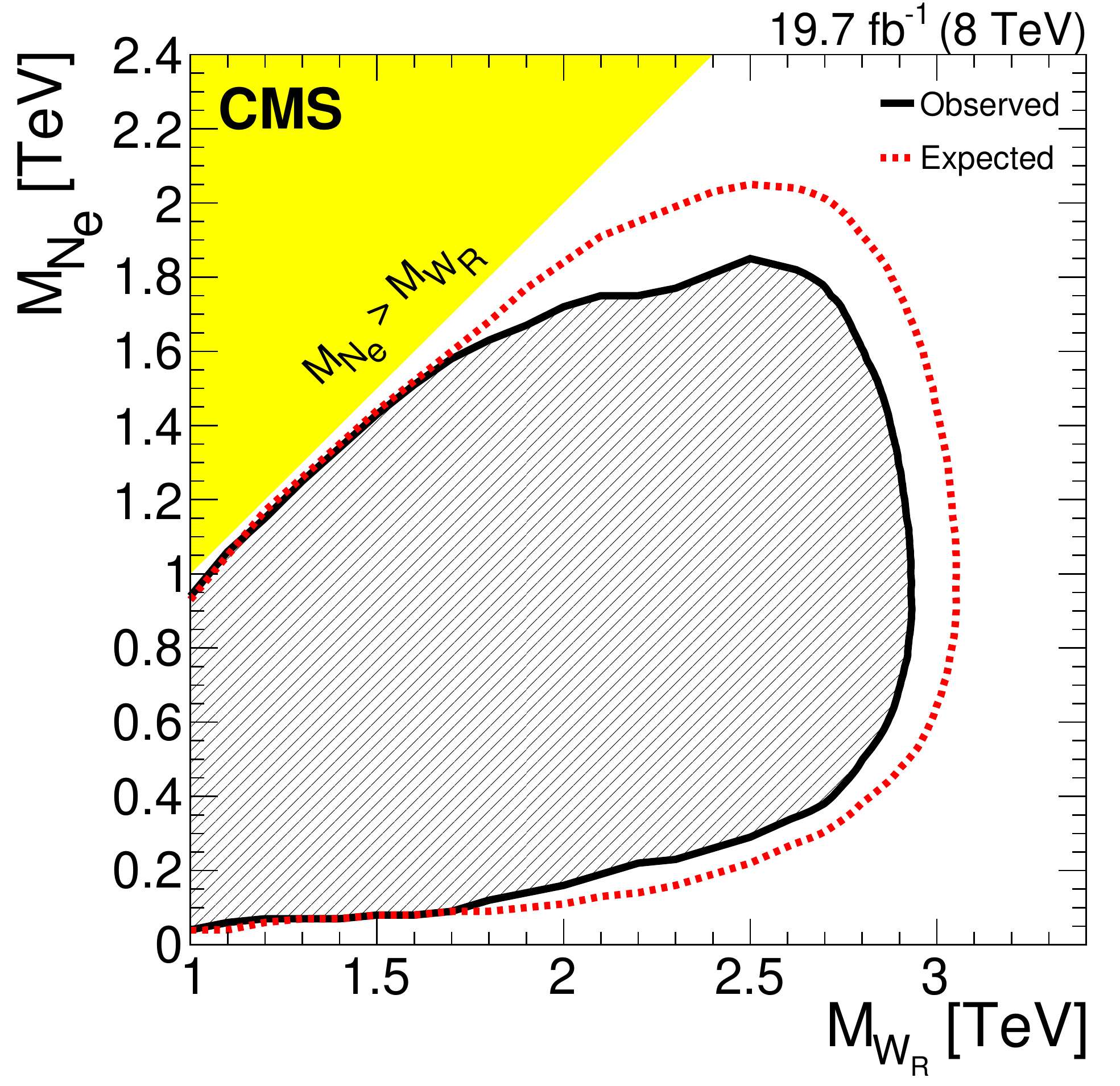}
\includegraphics[width=0.45\textwidth]{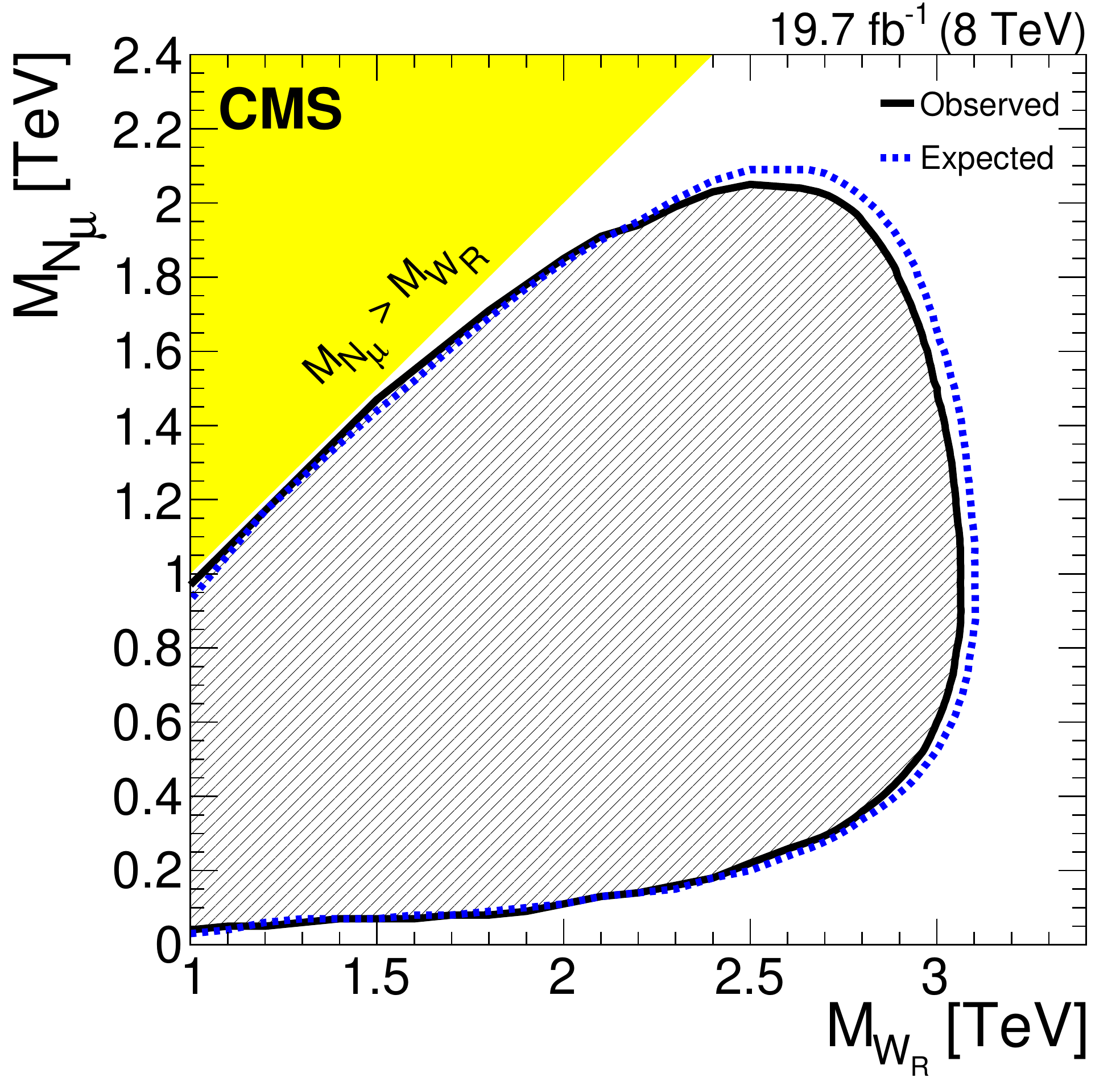}
\caption{The 95\% CL exclusion region (hatched) in the
$(M_{\WR},M_{\Nell})$ plane, assuming the model described in the text
(see Section~\ref{sec:intro}), for the electron (\cmsLeft) and muon (\cmsRight) channels.
Neutrino masses greater than \MWR (yellow shaded region) are not considered in this search.}
\label{fig:2dlimitsSolo}
\end{figure}

\begin{figure}[!htbp]
\includegraphics[width=0.45\textwidth]{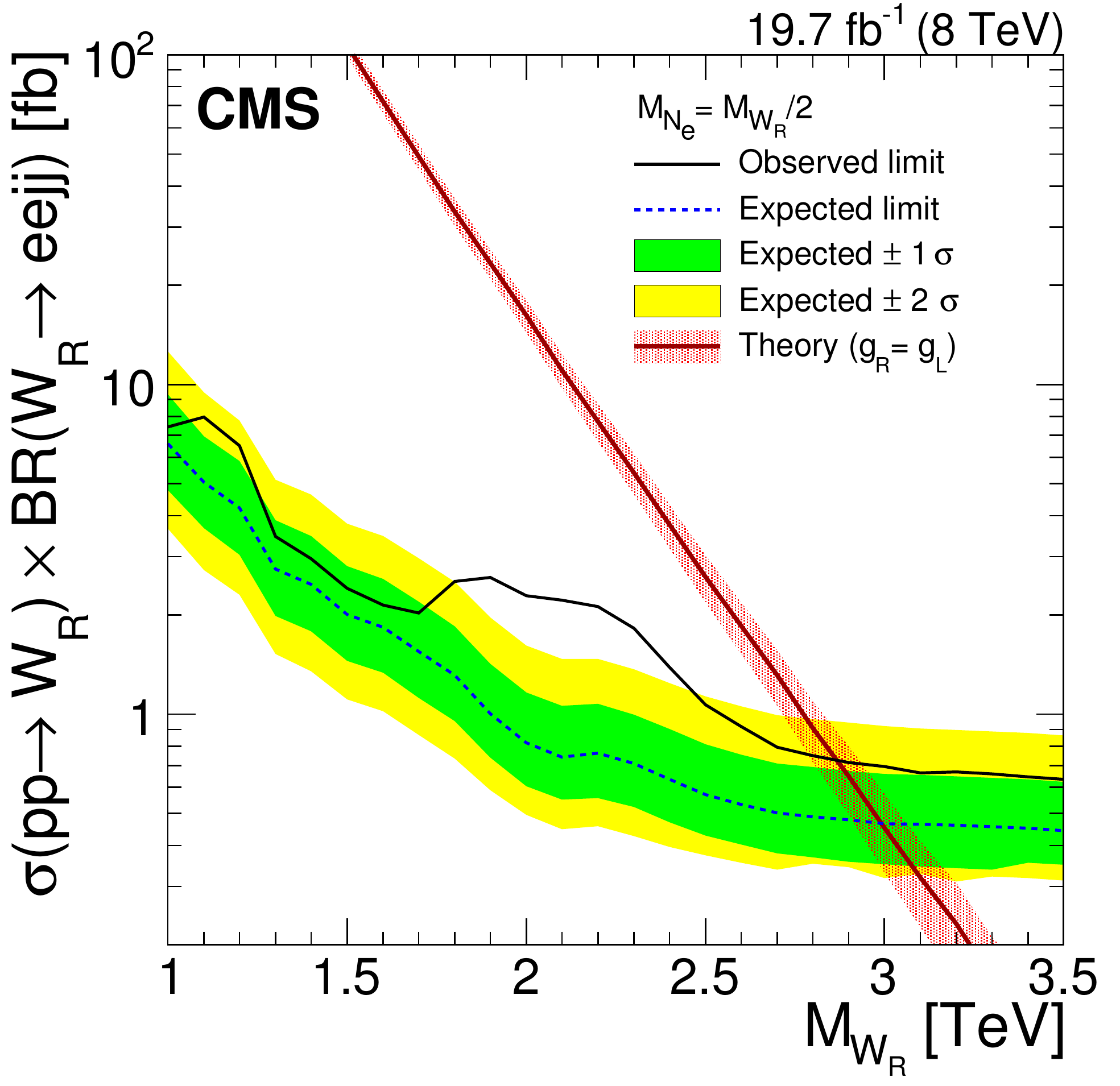}
\includegraphics[width=0.45\textwidth]{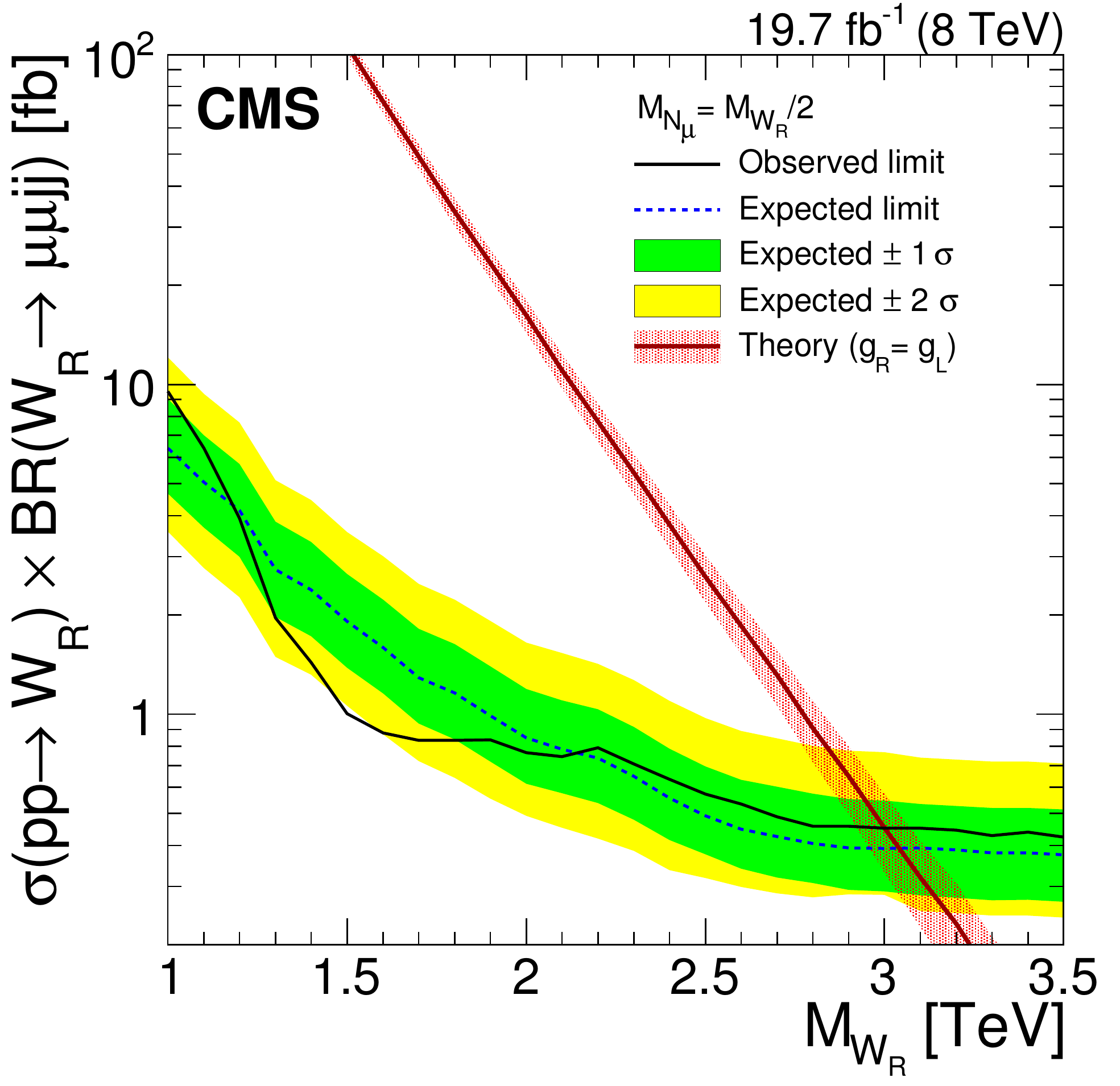}
\caption{The 95\% CL exclusion for \WR boson production cross section times branching
fraction, computed as a function of \MWR assuming the
right-handed neutrino has half the mass of the \WR boson, for the electron (\cmsLeft)
and muon (\cmsRight) channels.  The signal cross section PDF uncertainties (red band surrounding
the theoretical \WR-boson production cross section curve) are included for illustration purposes only.}
\label{fig:masslimitsWRsolo}
\end{figure}

For the results presented in Fig.~\ref{fig:2dlimitsSolo}, we indicate a range
of \Nell masses that are excluded as a function of \MWR assuming
that only one heavy neutrino flavor (electron or muon) is accessible from
8\TeV pp collisions, with the other $\cmsSymbolFace{N}_{\ell'}$
($\ell' = \Pe, \mu, \tau$, with $\ell' \neq \ell$) too heavy to be produced.
These $(M_{\WR}, M_{\Nell})$ limits are obtained by comparing the observed
and expected cross section upper limits with the expected cross section for
each mass point.  The limits extend to roughly $M_{\WR} = 3.0$\TeV in each channel
and exclude a wide range of heavy neutrino masses for \WR boson mass assumptions
below this maximal value.  The inclusion of the results from the
previous iteration of this analysis~\cite{EXO-11-091}, which searched
for \WR boson production in the $\mu \mu jj$ final state
using 7\TeV data, does not significantly affect
the limit results.  The excess in the electron channel at approximately
2\TeV has a local significance of 2.8$\sigma$
for a \WR boson candidate with a mass of 2.1\TeV.  Assuming contributions from SM
backgrounds only, the p-value for the local excess in the \Meejj distribution is 0.0050.
We also present limits as a function of \WR boson mass for a right-handed neutrino
with $M_{\Nell} = \frac{1}{2} M_{\WR}$ in Fig.~\ref{fig:masslimitsWRsolo}.  For the electron
(muon) channel, we exclude \WR bosons with $\MWR < 2.87\;(3.00)$\TeV, with an
expected exclusion of 2.99\;(3.04)\TeV.

\begin{figure}[!htbp]
\includegraphics[width=0.45\textwidth]{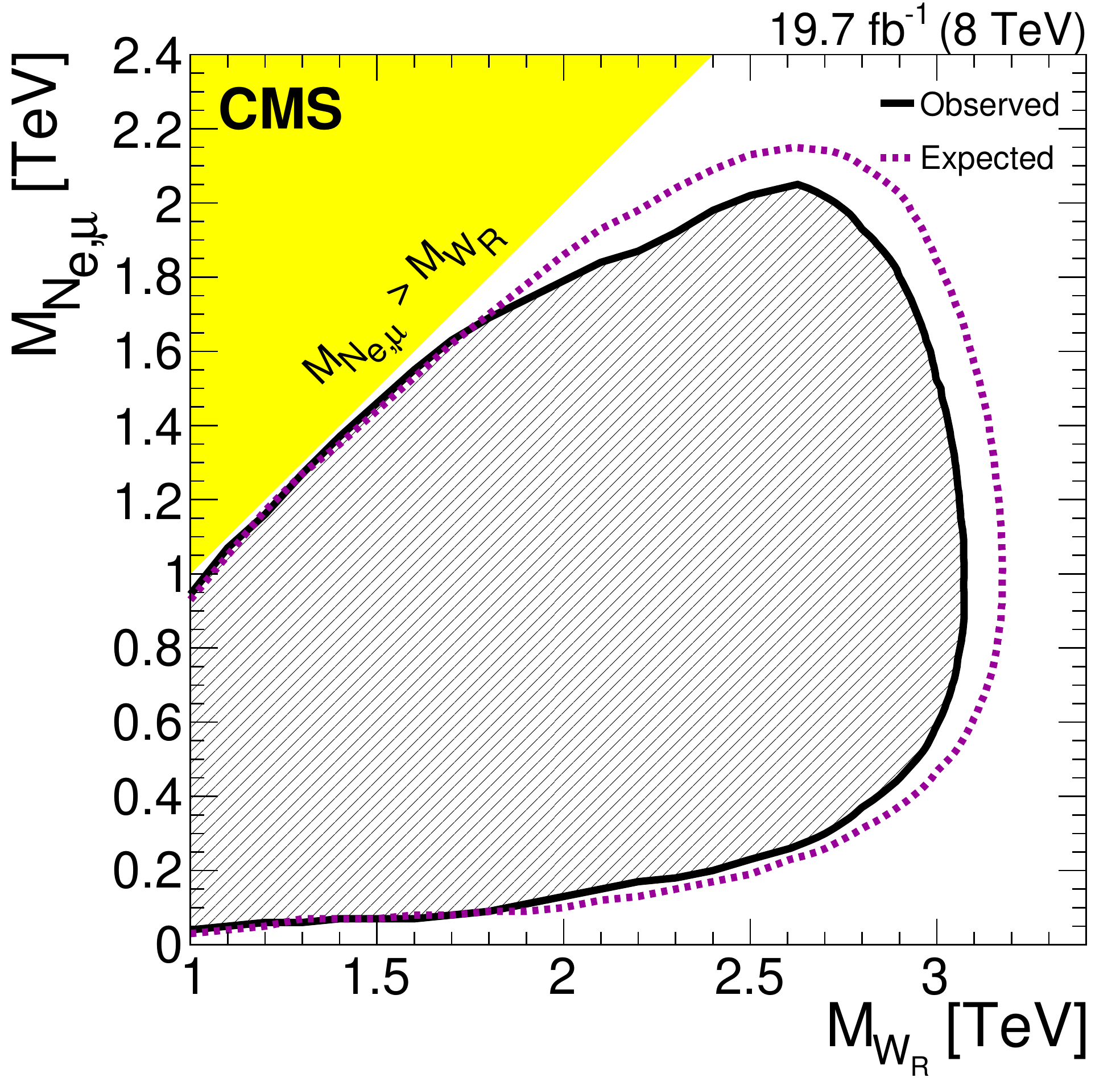}
\includegraphics[width=0.45\textwidth]{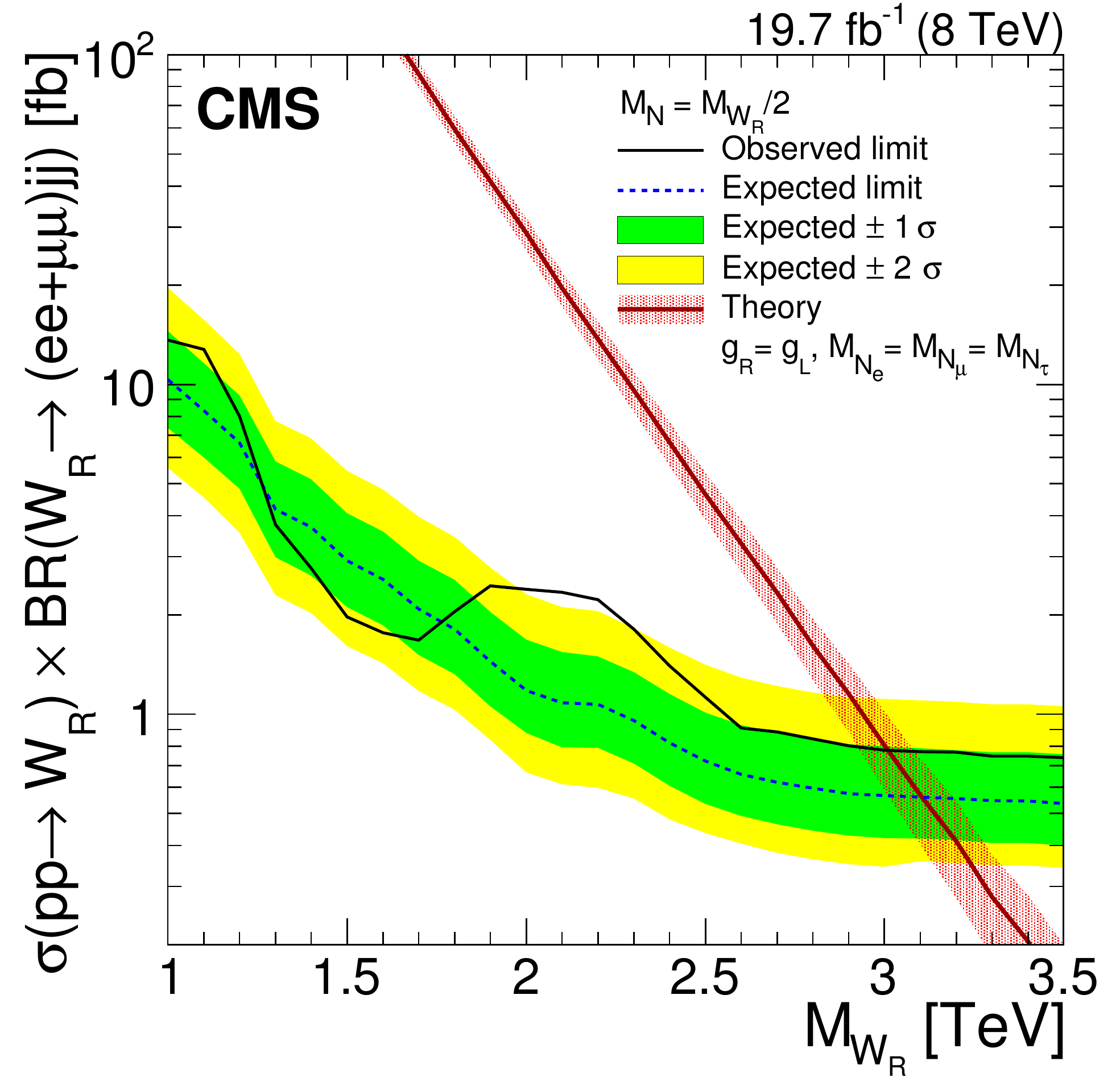}
\caption{The 95\% CL exclusion region in the
$(M_{\WR},M_{\Nell})$ plane (\cmsLeft), and as a function of \WR boson mass with
$M_N = \frac{1}{2} M_{\WR}$ (\cmsRight) obtained combining the electron
and muon channels.  The signal cross section PDF uncertainties (red band surrounding
the theoretical \WR-boson production cross section curve) are
included for illustration purposes only.  Neutrino masses greater
than \MWR (yellow shaded region in the \cmsLeft figure) are not considered in this search.}
\label{fig:limitsWRemu}
\end{figure}

We additionally consider the case where all \Nell masses are
degenerate and can be produced via \WR boson production and decay in 8\TeV pp collisions.
In this case, the electron and muon results can be combined as shown
in Fig.~\ref{fig:limitsWRemu}.  The $(\MWR,M_{\Nell})$ exclusion for the combination
extends slightly further than the single-channel exclusion limits, with an observed (expected)
exclusion for the combined channel of $\MWR < 3.01~(3.10)$\TeV for
$M_{\Nell} = \frac{1}{2} \MWR$.

\section{Summary}

A search for right-handed bosons (\WR) and heavy right-handed neutrinos (\Nell) in the
left-right symmetric extension of the standard model has been presented.
The data sample is in agreement with expectations from standard model
processes in the $\mu \mu j j$ final state.  An excess is observed in the electron channel
with a local significance of 2.8$\sigma$ at $\Meejj \approx 2.1$\TeV.
The excess does not appear to be consistent with expectations from left-right symmetric theory.
Considering $\WR \to \Pe \Ne$ and $\WR \to \mu \Nmu$ searches separately,
regions in the $(M_{\WR},M_{\Nell})$ mass space are excluded at 95\% confidence level
that extend up to
$M_{\WR} < 3.0$\TeV for both channels.  Assuming $\WR \to \ell \Nell$ with
degenerate \Nell mass for $\ell = \Pe, \mu$, \WR boson production is
excluded at 95\% confidence level up to $M_{\WR} < 3.0$\TeV.  This search has significantly
extended the exclusion region in the two-dimensional $(M_{\WR},M_{\Nell})$ mass plane compared to
previous searches, and for the first time this search has excluded \MWR values beyond
the theoretical lower mass limit of $\MWR \gtrsim 2.5$\TeV.

\section*{Acknowledgments}
{\tolerance=1200
We congratulate our colleagues in the CERN accelerator departments for the excellent performance of the LHC and thank the technical and administrative staffs at CERN and at other CMS institutes for their contributions to the success of the CMS effort. In addition, we gratefully acknowledge the computing centres and personnel of the Worldwide LHC Computing Grid for delivering so effectively the computing infrastructure essential to our analyses. Finally, we acknowledge the enduring support for the construction and operation of the LHC and the CMS detector provided by the following funding agencies: BMWFW and FWF (Austria); FNRS and FWO (Belgium); CNPq, CAPES, FAPERJ, and FAPESP (Brazil); MES (Bulgaria); CERN; CAS, MoST, and NSFC (China); COLCIENCIAS (Colombia); MSES and CSF (Croatia); RPF (Cyprus); MoER, ERC IUT and ERDF (Estonia); Academy of Finland, MEC, and HIP (Finland); CEA and CNRS/IN2P3 (France); BMBF, DFG, and HGF (Germany); GSRT (Greece); OTKA and NIH (Hungary); DAE and DST (India); IPM (Iran); SFI (Ireland); INFN (Italy); NRF and WCU (Republic of Korea); LAS (Lithuania); MOE and UM (Malaysia); CINVESTAV, CONACYT, SEP, and UASLP-FAI (Mexico); MBIE (New Zealand); PAEC (Pakistan); MSHE and NSC (Poland); FCT (Portugal); JINR (Dubna); MON, RosAtom, RAS and RFBR (Russia); MESTD (Serbia); SEIDI and CPAN (Spain); Swiss Funding Agencies (Switzerland); MST (Taipei); ThEPCenter, IPST, STAR and NSTDA (Thailand); TUBITAK and TAEK (Turkey); NASU and SFFR (Ukraine); STFC (United Kingdom); DOE and NSF (USA).

Individuals have received support from the Marie-Curie programme and the European Research Council and EPLANET (European Union); the Leventis Foundation; the A. P. Sloan Foundation; the Alexander von Humboldt Foundation; the Belgian Federal Science Policy Office; the Fonds pour la Formation \`a la Recherche dans l'Industrie et dans l'Agriculture (FRIA-Belgium); the Agentschap voor Innovatie door Wetenschap en Technologie (IWT-Belgium); the Ministry of Education, Youth and Sports (MEYS) of the Czech Republic; the Council of Science and Industrial Research, India; the HOMING PLUS programme of Foundation for Polish Science, cofinanced from European Union, Regional Development Fund; the Compagnia di San Paolo (Torino); the Consorzio per la Fisica (Trieste); MIUR project 20108T4XTM (Italy); the Thalis and Aristeia programmes cofinanced by EU-ESF and the Greek NSRF; and the National Priorities Research Program by Qatar National Research Fund.
\par}

\bibliography{auto_generated}   

\ifthenelse{\boolean{cms@external}}{}{
\clearpage
\appendix
\setcounter{table}{0}\renewcommand{\thetable}{\thesection\arabic{table}}
\section{95\% CL exclusion limits as a function of \texorpdfstring{\WR and \Nell}{W[R] and N[R]} mass (tabular format)\label{app:suppMat}}
\begin{table*}[!htbp]
\centering
\topcaption{The 95\% CL observed (Obs.) and expected (Exp.) exclusion limits (in fb) on the \WR production cross section times branching fraction for $\WR \to \Pe \Pe j j$ as a function of \WR and \Ne mass (in \GeVns) for $1000 \leq \MWR \leq 1600$\GeV.  The signal acceptance (Acc.) is also included for each $(\MWR,M_{\Ne})$ entry.}
\begin{tabular}{ccccc|ccccc}
\hline
\MWR & $M_{\Ne}$ & Obs. & Exp. & Acc. & \MWR &  $M_{\Ne}$ & Obs. & Exp. & Acc. \\
\hline
1000  &   100  &  65.7  &  58.5  &  0.073 $\pm$ 0.110  &  1400  &   100  &  60.1  &  50.2  &  0.043 $\pm$ 0.164  \\
1000  &   200  &  14.6  &  13.0  &  0.298 $\pm$ 0.031  &  1400  &   200  &  9.07  &  7.57  &  0.241 $\pm$ 0.061  \\
1000  &   300  &  9.77  &  8.69  &  0.447 $\pm$ 0.014  &  1400  &   300  &  4.91  &  4.09  &  0.428 $\pm$ 0.027  \\
1000  &   400  &  8.17  &  7.27  &  0.549 $\pm$ 0.011  &  1400  &   400  &  3.76  &  3.13  &  0.549 $\pm$ 0.015  \\
1000  &   500  &  7.44  &  6.63  &  0.596 $\pm$ 0.009  &  1400  &   500  &  3.29  &  2.75  &  0.630 $\pm$ 0.011  \\
1000  &   600  &  7.14  &  6.35  &  0.620 $\pm$ 0.009  &  1400  &   600  &  3.09  &  2.58  &  0.688 $\pm$ 0.010  \\
1000  &   700  &  7.16  &  6.38  &  0.617 $\pm$ 0.009  &  1400  &   700  &  2.97  &  2.48  &  0.711 $\pm$ 0.011  \\
\hline
\end{tabular}
\label{tab:eLimits1000to1600}
\end{table*}

\begin{table*}[!htbp]
\centering
\topcaption{The 95\% CL observed (Obs.) and expected (Exp.) exclusion limits (in fb) on the \WR production cross section times branching fraction for $\WR \to \Pe \Pe j j$ as a function of \WR and \Ne mass (in \GeVns) for $1700 \leq \MWR \leq 2000$\GeV.  The signal acceptance (Acc.) is also included for each $(\MWR,M_{\Ne})$ entry.}
\begin{tabular}{ccccc|ccccc}
\hline
\MWR & $M_{\Ne}$ & Obs. & Exp. & Acc. & \MWR &  $M_{\Ne}$ & Obs. & Exp. & Acc. \\
\hline
1700  &   100  &  76.5  &  58.3  &  0.032 $\pm$ 0.197  &  1900  &   100  &  131  &  50.7  &  0.025 $\pm$ 0.216  \\
1700  &   200  &  9.12  &  6.95  &  0.195 $\pm$ 0.086  &  1900  &   200  &  14.3  &  5.51  &  0.161 $\pm$ 0.103  \\
1700  &   300  &  3.87  &  2.95  &  0.383 $\pm$ 0.040  &  1900  &   300  &  5.83  &  2.25  &  0.348 $\pm$ 0.055  \\
1700  &   400  &  2.71  &  2.07  &  0.526 $\pm$ 0.022  &  1900  &   400  &  3.91  &  1.51  &  0.502 $\pm$ 0.028  \\
1700  &   500  &  2.30  &  1.75  &  0.616 $\pm$ 0.015  &  1900  &   500  &  3.23  &  1.25  &  0.605 $\pm$ 0.025  \\
1700  &   600  &  2.14  &  1.63  &  0.676 $\pm$ 0.011  &  1900  &   600  &  2.89  &  1.12  &  0.674 $\pm$ 0.013  \\
1700  &   700  &  2.09  &  1.59  &  0.719 $\pm$ 0.010  &  1900  &   700  &  2.74  &  1.06  &  0.715 $\pm$ 0.011  \\
\hline
\end{tabular}
\label{tab:eLimits1700to2000}
\end{table*}

\begin{table*}[!htbp]
\centering
\topcaption{The 95\% CL observed (Obs.) and expected (Exp.) exclusion limits (in fb) on the \WR production cross section times branching fraction for $\WR \to \Pe \Pe j j$ as a function of \WR and \Ne mass (in \GeVns) for $2100 \leq \MWR \leq 2300$\GeV.  The signal acceptance (Acc.) is also included for each $(\MWR,M_{\Ne})$ entry.}
\begin{tabular}{ccccc|ccccc}
\hline
\MWR & $M_{\Ne}$ & Obs. & Exp. & Acc. & \MWR &  $M_{\Ne}$ & Obs. & Exp. & Acc. \\
\hline
2100  &   100  &  173  &  57.8  &  0.024 $\pm$ 0.233  &  2300  &   100  &  158  &  61.5  &  0.022 $\pm$ 0.251  \\
2100  &   200  &  18.1  &  6.07  &  0.142 $\pm$ 0.117  &  2300  &   200  &  15.5  &  6.00  &  0.134 $\pm$ 0.133  \\
2100  &   300  &  6.94  &  2.32  &  0.315 $\pm$ 0.061  &  2300  &   300  &  5.76  &  2.24  &  0.297 $\pm$ 0.072  \\
2100  &   400  &  4.26  &  1.42  &  0.469 $\pm$ 0.034  &  2300  &   400  &  3.42  &  1.33  &  0.438 $\pm$ 0.042  \\
2100  &   500  &  3.25  &  1.09  &  0.586 $\pm$ 0.021  &  2300  &   500  &  2.62  &  1.02  &  0.554 $\pm$ 0.026  \\
2100  &   600  &  2.77  &  0.927  &  0.657 $\pm$ 0.015  &  2300  &   600  &  2.26  &  0.879  &  0.651 $\pm$ 0.018  \\
2100  &   700  &  2.52  &  0.843  &  0.710 $\pm$ 0.012  &  2300  &   700  &  2.06  &  0.798  &  0.691 $\pm$ 0.014  \\
\hline
\end{tabular}
\label{tab:eLimits2100to2300}
\end{table*}

\begin{table*}[!htbp]
\centering
\topcaption{The 95\% CL observed (Obs.) and expected (Exp.) exclusion limits (in fb) on the \WR production cross section times branching fraction for $\WR \to \Pe \Pe j j$ as a function of \WR and \Ne mass (in \GeVns) for $2400 \leq \MWR \leq 2500$\GeV.  The signal acceptance (Acc.) is also included for each $(\MWR,M_{\Ne})$ entry.}
\begin{tabular}{ccccc|ccccc}
\hline
\MWR & $M_{\Ne}$ & Obs. & Exp. & Acc. & \MWR &  $M_{\Ne}$ & Obs. & Exp. & Acc. \\
\hline
2400  &   100  &  160  &  73.2  &  0.025 $\pm$ 0.262  &  2500  &   100  &  157  &  83.8  &  0.027 $\pm$ 0.270  \\
2400  &   200  &  14.3  &  6.57  &  0.127 $\pm$ 0.139  &  2500  &   200  &  13.6  &  7.27  &  0.125 $\pm$ 0.149  \\
2400  &   300  &  5.09  &  2.33  &  0.278 $\pm$ 0.078  &  2500  &   300  &  4.63  &  2.47  &  0.265 $\pm$ 0.085  \\
2400  &   400  &  2.93  &  1.34  &  0.429 $\pm$ 0.060  &  2500  &   400  &  2.60  &  1.39  &  0.414 $\pm$ 0.059  \\
2400  &   500  &  2.16  &  0.990  &  0.548 $\pm$ 0.028  &  2500  &   500  &  1.82  &  0.972  &  0.539 $\pm$ 0.032  \\
2400  &   600  &  1.79  &  0.821  &  0.634 $\pm$ 0.019  &  2500  &   600  &  1.50  &  0.800  &  0.622 $\pm$ 0.021  \\
2400  &   700  &  1.63  &  0.745  &  0.691 $\pm$ 0.015  &  2500  &   700  &  1.32  &  0.705  &  0.683 $\pm$ 0.015  \\
\hline
\end{tabular}
\label{tab:eLimits2400to2500}
\end{table*}

\begin{table*}[!htbp]
\centering
\topcaption{The 95\% CL observed (Obs.) and expected (Exp.) exclusion limits (in fb) on the \WR production cross section times branching fraction for $\WR \to \Pe \Pe j j$ as a function of \WR and \Ne mass (in \GeVns) for $2600 \leq \MWR \leq 2700$\GeV.  The signal acceptance (Acc.) is also included for each $(\MWR,M_{\Ne})$ entry.}
\begin{tabular}{ccccc|ccccc}
\hline
\MWR & $M_{\Ne}$ & Obs. & Exp. & Acc. & \MWR &  $M_{\Ne}$ & Obs. & Exp. & Acc. \\
\hline
2600  &   100  &  148  &  86.1  &  0.023 $\pm$ 0.276  &  2700  &   100  &  131  &  83.0  &  0.026 $\pm$ 0.288  \\
2600  &   200  &  13.9  &  8.07  &  0.122 $\pm$ 0.157  &  2700  &   200  &  13.2  &  8.35  &  0.127 $\pm$ 0.165  \\
2600  &   300  &  4.60  &  2.67  &  0.256 $\pm$ 0.090  &  2700  &   300  &  4.36  &  2.76  &  0.257 $\pm$ 0.098  \\
2600  &   400  &  2.46  &  1.43  &  0.404 $\pm$ 0.054  &  2700  &   400  &  2.29  &  1.45  &  0.394 $\pm$ 0.059  \\
2600  &   500  &  1.71  &  0.993  &  0.521 $\pm$ 0.034  &  2700  &   500  &  1.56  &  0.986  &  0.513 $\pm$ 0.037  \\
2600  &   600  &  1.37  &  0.794  &  0.616 $\pm$ 0.024  &  2700  &   600  &  1.22  &  0.770  &  0.613 $\pm$ 0.025  \\
2600  &   700  &  1.19  &  0.689  &  0.682 $\pm$ 0.018  &  2700  &   700  &  1.07  &  0.678  &  0.674 $\pm$ 0.018  \\
\hline
\end{tabular}
\label{tab:eLimits2600to2700}
\end{table*}

\begin{table*}[!htbp]
\centering
\topcaption{The 95\% CL observed (Obs.) and expected (Exp.) exclusion limits (in fb) on the \WR production cross section times branching fraction for $\WR \to \Pe \Pe j j$ as a function of \WR and \Ne mass (in \GeVns) for $2800 \leq \MWR \leq 2900$\GeV.  The signal acceptance (Acc.) is also included for each $(\MWR,M_{\Ne})$ entry.}
\begin{tabular}{ccccc|ccccc}
\hline
\MWR & $M_{\Ne}$ & Obs. & Exp. & Acc. & \MWR &  $M_{\Ne}$ & Obs. & Exp. & Acc. \\
\hline
2800  &   100  &  138  &  90.4  &  0.028 $\pm$ 0.300  &  2900  &   100  &  136  &  91.3  &  0.030 $\pm$ 0.307  \\
2800  &   200  &  15.2  &  9.93  &  0.130 $\pm$ 0.175  &  2900  &   200  &  16.8  &  11.2  &  0.130 $\pm$ 0.187  \\
2800  &   300  &  5.01  &  3.27  &  0.255 $\pm$ 0.108  &  2900  &   300  &  5.42  &  3.62  &  0.247 $\pm$ 0.115  \\
2800  &   400  &  2.56  &  1.67  &  0.385 $\pm$ 0.064  &  2900  &   400  &  2.71  &  1.81  &  0.391 $\pm$ 0.071  \\
2800  &   500  &  1.69  &  1.10  &  0.507 $\pm$ 0.041  &  2900  &   500  &  1.74  &  1.16  &  0.498 $\pm$ 0.045  \\
2800  &   600  &  1.30  &  0.851  &  0.601 $\pm$ 0.027  &  2900  &   600  &  1.32  &  0.880  &  0.595 $\pm$ 0.029  \\
2800  &   700  &  1.09  &  0.713  &  0.666 $\pm$ 0.020  &  2900  &   700  &  1.10  &  0.735  &  0.658 $\pm$ 0.022  \\
\hline
\end{tabular}
\label{tab:eLimits2800to2900}
\end{table*}

\clearpage

\begin{table*}[!htbp]
\centering
\topcaption{The 95\% CL observed (Obs.) and expected (Exp.) exclusion limits (in fb) on the \WR production cross section times branching fraction for $\WR \to \Pe \Pe j j$ as a function of \WR and \Ne mass (in \GeVns) for $3000 \leq \MWR \leq 3100$\GeV.  The signal acceptance (Acc.) is also included for each $(\MWR,M_{\Ne})$ entry.}
\begin{tabular}{ccccc|ccccc}
\hline
\MWR & $M_{\Ne}$ & Obs. & Exp. & Acc. & \MWR &  $M_{\Ne}$ & Obs. & Exp. & Acc. \\
\hline
3000  &   100  &  134  &  90.2  &  0.030 $\pm$ 0.314  &  3100  &   100  &  134  &  94.2  &  0.036 $\pm$ 0.318  \\
3000  &   200  &  18.8  &  12.6  &  0.131 $\pm$ 0.196  &  3100  &   200  &  20.6  &  14.4  &  0.145 $\pm$ 0.204  \\
3000  &   300  &  5.90  &  3.96  &  0.256 $\pm$ 0.124  &  3100  &   300  &  6.52  &  4.57  &  0.261 $\pm$ 0.136  \\
3000  &   400  &  2.94  &  1.97  &  0.382 $\pm$ 0.082  &  3100  &   400  &  3.12  &  2.19  &  0.387 $\pm$ 0.089  \\
3000  &   500  &  1.86  &  1.25  &  0.496 $\pm$ 0.052  &  3100  &   500  &  1.97  &  1.38  &  0.489 $\pm$ 0.057  \\
3000  &   600  &  1.37  &  0.923  &  0.586 $\pm$ 0.033  &  3100  &   600  &  1.42  &  0.996  &  0.578 $\pm$ 0.037  \\
3000  &   700  &  1.13  &  0.756  &  0.648 $\pm$ 0.022  &  3100  &   700  &  1.15  &  0.806  &  0.646 $\pm$ 0.028  \\
\hline
\end{tabular}
\label{tab:eLimits3000to3100}
\end{table*}

\begin{table*}[!htbp]
\centering
\topcaption{The 95\% CL observed (Obs.) and expected (Exp.) exclusion limits (in fb) on the \WR production cross section times branching fraction for $\WR \to \Pe \Pe j j$ as a function of \WR and \Ne mass (in \GeVns) for $ \MWR = 3200$\GeV.  The signal acceptance (Acc.) is also included for each $(\MWR,M_{\Ne})$ entry.}
\begin{tabular}{ccccc}
\hline
\MWR & $M_{\Ne}$ & Obs. & Exp. & Acc. \\
\hline
3200  &   100  &  139  &  95.7  &  0.039 $\pm$ 0.320  \\
3200  &   200  &  23.5  &  16.3  &  0.147 $\pm$ 0.210  \\
3200  &   300  &  7.35  &  5.08  &  0.277 $\pm$ 0.143  \\
3200  &   400  &  3.56  &  2.46  &  0.391 $\pm$ 0.096  \\
3200  &   500  &  2.19  &  1.51  &  0.485 $\pm$ 0.066  \\
3200  &   600  &  1.57  &  1.08  &  0.574 $\pm$ 0.041  \\
3200  &   700  &  1.25  &  0.861  &  0.639 $\pm$ 0.032  \\
3200  &   800  &  1.06  &  0.731  &  0.690 $\pm$ 0.022  \\
3200  &   900  &  0.936  &  0.646  &  0.721 $\pm$ 0.025  \\
3200  &  1000  &  0.860  &  0.594  &  0.749 $\pm$ 0.019  \\
3200  &  1100  &  0.804  &  0.555  &  0.774 $\pm$ 0.016  \\
3200  &  1200  &  0.760  &  0.524  &  0.787 $\pm$ 0.012  \\
3200  &  1300  &  0.731  &  0.504  &  0.800 $\pm$ 0.017  \\
3200  &  1400  &  0.705  &  0.487  &  0.809 $\pm$ 0.021  \\
3200  &  1500  &  0.684  &  0.472  &  0.818 $\pm$ 0.014  \\
3200  &  1600  &  0.669  &  0.462  &  0.827 $\pm$ 0.014  \\
3200  &  1700  &  0.667  &  0.461  &  0.828 $\pm$ 0.011  \\
3200  &  1800  &  0.662  &  0.457  &  0.834 $\pm$ 0.019  \\
3200  &  1900  &  0.661  &  0.456  &  0.835 $\pm$ 0.011  \\
3200  &  2000  &  0.659  &  0.455  &  0.837 $\pm$ 0.033  \\
3200  &  2100  &  0.664  &  0.459  &  0.831 $\pm$ 0.010  \\
3200  &  2200  &  0.658  &  0.454  &  0.838 $\pm$ 0.018  \\
3200  &  2300  &  0.662  &  0.457  &  0.833 $\pm$ 0.011  \\
3200  &  2400  &  0.666  &  0.460  &  0.828 $\pm$ 0.010  \\
3200  &  2500  &  0.671  &  0.463  &  0.822 $\pm$ 0.011  \\
3200  &  2600  &  0.676  &  0.467  &  0.816 $\pm$ 0.010  \\
3200  &  2700  &  0.678  &  0.468  &  0.813 $\pm$ 0.015  \\
3200  &  2800  &  0.690  &  0.476  &  0.800 $\pm$ 0.010  \\
3200  &  2900  &  0.709  &  0.489  &  0.778 $\pm$ 0.013  \\
3200  &  3000  &  0.730  &  0.504  &  0.755 $\pm$ 0.012  \\
3200  &  3100  &  0.794  &  0.548  &  0.695 $\pm$ 0.028  \\
\hline
\end{tabular}
\label{tab:eLimits3200to3200}
\end{table*}

\begin{table*}[!htbp]
\centering
\topcaption{The 95\% CL observed (Obs.) and expected (Exp.) exclusion limits (in fb) on the \WR production cross section times branching fraction for $\WR \to \mu \mu j j$ as a function of \WR and \Nmu mass (in \GeVns) for $1000 \leq \MWR \leq 1600$\GeV.  This signal acceptance (Acc.) is also included for each $(\MWR,M_{\Nmu})$ entry.}
\begin{tabular}{ccccc|ccccc}
\hline
\MWR & $M_{\Nmu}$ & Obs. & Exp. & Acc. & \MWR &  $M_{\Nmu}$ & Obs. & Exp. & Acc. \\
\hline
1000  &   100  &  79.0  &  53.1  &  0.0664 $\pm$ 0.0685  &  1400  &   100  &  26.9  &  44.7  &  0.037 $\pm$ 0.140  \\
1000  &   200  &  18.1  &  12.1  &  0.283 $\pm$ 0.017  &  1400  &   200  &  4.17  &  6.93  &  0.224 $\pm$ 0.029  \\
1000  &   300  &  12.4  &  8.31  &  0.435 $\pm$ 0.014  &  1400  &   300  &  2.31  &  3.83  &  0.410 $\pm$ 0.016  \\
1000  &   400  &  10.4  &  7.00  &  0.539 $\pm$ 0.014  &  1400  &   400  &  1.79  &  2.97  &  0.533 $\pm$ 0.015  \\
1000  &   500  &  9.53  &  6.40  &  0.586 $\pm$ 0.015  &  1400  &   500  &  1.58  &  2.63  &  0.618 $\pm$ 0.015  \\
1000  &   600  &  9.12  &  6.13  &  0.612 $\pm$ 0.014  &  1400  &   600  &  1.49  &  2.48  &  0.676 $\pm$ 0.014  \\
1000  &   700  &  9.16  &  6.15  &  0.609 $\pm$ 0.014  &  1400  &   700  &  1.43  &  2.38  &  0.702 $\pm$ 0.014  \\
\hline
\end{tabular}
\label{tab:muLimits1000to1600}
\end{table*}

\begin{table*}[!htbp]
\centering
\topcaption{The 95\% CL observed (Obs.) and expected (Exp.) exclusion limits (in fb) on the \WR production cross section times branching fraction for $\WR \to \mu \mu j j$ as a function of \WR and \Nmu mass (in \GeVns) for $1700 \leq \MWR \leq 2000$\GeV.  This signal acceptance (Acc.) is also included for each $(\MWR,M_{\Nmu})$ entry.}
\begin{tabular}{ccccc|ccccc}
\hline
\MWR & $M_{\Nmu}$ & Obs. & Exp. & Acc. & \MWR &  $M_{\Nmu}$ & Obs. & Exp. & Acc. \\
\hline
1700  &   100  &  28.6  &  44.3  &  0.028 $\pm$ 0.198  &  1900  &   100  &  38.2  &  45.2  &  0.020 $\pm$ 0.235  \\
1700  &   200  &  3.53  &  5.46  &  0.179 $\pm$ 0.048  &  1900  &   200  &  4.32  &  5.11  &  0.148 $\pm$ 0.063  \\
1700  &   300  &  1.53  &  2.36  &  0.364 $\pm$ 0.020  &  1900  &   300  &  1.80  &  2.12  &  0.327 $\pm$ 0.028  \\
1700  &   400  &  1.09  &  1.68  &  0.507 $\pm$ 0.016  &  1900  &   400  &  1.22  &  1.45  &  0.482 $\pm$ 0.023  \\
1700  &   500  &  0.929  &  1.44  &  0.601 $\pm$ 0.014  &  1900  &   500  &  1.02  &  1.21  &  0.587 $\pm$ 0.014  \\
1700  &   600  &  0.869  &  1.35  &  0.664 $\pm$ 0.015  &  1900  &   600  &  0.922  &  1.09  &  0.659 $\pm$ 0.016  \\
1700  &   700  &  0.853  &  1.32  &  0.709 $\pm$ 0.014  &  1900  &   700  &  0.879  &  1.04  &  0.703 $\pm$ 0.015  \\
\hline
\end{tabular}
\label{tab:muLimits1700to2000}
\end{table*}

\begin{table*}[!htbp]
\centering
\topcaption{The 95\% CL observed (Obs.) and expected (Exp.) exclusion limits (in fb) on the \WR production cross section times branching fraction for $\WR \to \mu \mu j j$ as a function of \WR and \Nmu mass (in \GeVns) for $2100 \leq \MWR \leq 2300$\GeV.  This signal acceptance (Acc.) is also included for each $(\MWR,M_{\Nmu})$ entry.}
\begin{tabular}{ccccc|ccccc}
\hline
\MWR & $M_{\Nmu}$ & Obs. & Exp. & Acc. & \MWR &  $M_{\Nmu}$ & Obs. & Exp. & Acc. \\
\hline
2100  &   100  &  51.7  &  54.3  &  0.020 $\pm$ 0.271  &  2300  &   100  &  54.0  &  49.5  &  0.018 $\pm$ 0.306  \\
2100  &   200  &  5.70  &  5.99  &  0.129 $\pm$ 0.076  &  2300  &   200  &  5.59  &  5.12  &  0.121 $\pm$ 0.095  \\
2100  &   300  &  2.22  &  2.33  &  0.294 $\pm$ 0.029  &  2300  &   300  &  2.11  &  1.94  &  0.275 $\pm$ 0.037  \\
2100  &   400  &  1.38  &  1.45  &  0.448 $\pm$ 0.018  &  2300  &   400  &  1.27  &  1.17  &  0.413 $\pm$ 0.020  \\
2100  &   500  &  1.07  &  1.12  &  0.566 $\pm$ 0.025  &  2300  &   500  &  0.987  &  0.905  &  0.533 $\pm$ 0.016  \\
2100  &   600  &  0.917  &  0.964  &  0.640 $\pm$ 0.014  &  2300  &   600  &  0.862  &  0.791  &  0.632 $\pm$ 0.015  \\
2100  &   700  &  0.839  &  0.881  &  0.696 $\pm$ 0.014  &  2300  &   700  &  0.788  &  0.722  &  0.675 $\pm$ 0.014  \\
\hline
\end{tabular}
\label{tab:muLimits2100to2300}
\end{table*}

\begin{table*}[!htbp]
\centering
\topcaption{The 95\% CL observed (Obs.) and expected (Exp.) exclusion limits (in fb) on the \WR production cross section times branching fraction for $\WR \to \mu \mu j j$ as a function of \WR and \Nmu mass (in \GeVns) for $2400 \leq \MWR \leq 2500$\GeV.  This signal acceptance (Acc.) is also included for each $(\MWR,M_{\Nmu})$ entry.}
\begin{tabular}{ccccc|ccccc}
\hline
\MWR & $M_{\Nmu}$ & Obs. & Exp. & Acc. & \MWR &  $M_{\Nmu}$ & Obs. & Exp. & Acc. \\
\hline
2400  &   100  &  63.9  &  56.1  &  0.020 $\pm$ 0.325  &  2500  &   100  &  72.8  &  62.7  &  0.021 $\pm$ 0.345  \\
2400  &   200  &  6.12  &  5.36  &  0.116 $\pm$ 0.105  &  2500  &   200  &  6.78  &  5.83  &  0.112 $\pm$ 0.116  \\
2400  &   300  &  2.20  &  1.93  &  0.258 $\pm$ 0.041  &  2500  &   300  &  2.33  &  2.01  &  0.245 $\pm$ 0.047  \\
2400  &   400  &  1.29  &  1.13  &  0.407 $\pm$ 0.022  &  2500  &   400  &  1.33  &  1.14  &  0.389 $\pm$ 0.024  \\
2400  &   500  &  0.961  &  0.843  &  0.526 $\pm$ 0.016  &  2500  &   500  &  0.944  &  0.813  &  0.515 $\pm$ 0.017  \\
2400  &   600  &  0.804  &  0.705  &  0.614 $\pm$ 0.015  &  2500  &   600  &  0.784  &  0.675  &  0.602 $\pm$ 0.016  \\
2400  &   700  &  0.735  &  0.645  &  0.675 $\pm$ 0.014  &  2500  &   700  &  0.697  &  0.600  &  0.665 $\pm$ 0.014  \\
\hline
\end{tabular}
\label{tab:muLimits2400to2500}
\end{table*}

\begin{table*}[!htbp]
\centering
\topcaption{The 95\% CL observed (Obs.) and expected (Exp.) exclusion limits (in fb) on the \WR production cross section times branching fraction for $\WR \to \mu \mu j j$ as a function of \WR and \Nmu mass (in \GeVns) for $2600 \leq \MWR \leq 2700$\GeV.  This signal acceptance (Acc.) is also included for each $(\MWR,M_{\Nmu})$ entry.}
\begin{tabular}{ccccc|ccccc}
\hline
\MWR & $M_{\Nmu}$ & Obs. & Exp. & Acc. & \MWR &  $M_{\Nmu}$ & Obs. & Exp. & Acc. \\
\hline
2600  &   100  &  74.3  &  62.3  &  0.019 $\pm$ 0.359  &  2700  &   100  &  68.8  &  59.9  &  0.021 $\pm$ 0.378  \\
2600  &   200  &  7.52  &  6.30  &  0.110 $\pm$ 0.127  &  2700  &   200  &  7.51  &  6.55  &  0.114 $\pm$ 0.142  \\
2600  &   300  &  2.53  &  2.12  &  0.236 $\pm$ 0.053  &  2700  &   300  &  2.52  &  2.20  &  0.237 $\pm$ 0.060  \\
2600  &   400  &  1.37  &  1.15  &  0.379 $\pm$ 0.029  &  2700  &   400  &  1.34  &  1.17  &  0.369 $\pm$ 0.032  \\
2600  &   500  &  0.963  &  0.807  &  0.498 $\pm$ 0.018  &  2700  &   500  &  0.923  &  0.805  &  0.487 $\pm$ 0.019  \\
2600  &   600  &  0.777  &  0.651  &  0.594 $\pm$ 0.015  &  2700  &   600  &  0.728  &  0.635  &  0.589 $\pm$ 0.015  \\
2600  &   700  &  0.680  &  0.570  &  0.663 $\pm$ 0.014  &  2700  &   700  &  0.646  &  0.564  &  0.652 $\pm$ 0.015  \\
\hline
\end{tabular}
\label{tab:muLimits2600to2700}
\end{table*}

\begin{table*}[!htbp]
\centering
\topcaption{The 95\% CL observed (Obs.) and expected (Exp.) exclusion limits (in fb) on the \WR production cross section times branching fraction for $\WR \to \mu \mu j j$ as a function of \WR and \Nmu mass (in \GeVns) for $2800 \leq \MWR \leq 2900$\GeV.  This signal acceptance (Acc.) is also included for each $(\MWR,M_{\Nmu})$ entry.}
\begin{tabular}{ccccc|ccccc}
\hline
\MWR & $M_{\Nmu}$ & Obs. & Exp. & Acc. & \MWR &  $M_{\Nmu}$ & Obs. & Exp. & Acc. \\
\hline
2800  &   100  &  71.6  &  63.4  &  0.023 $\pm$ 0.398  &  2900  &   100  &  73.4  &  63.2  &  0.024 $\pm$ 0.412  \\
2800  &   200  &  8.59  &  7.62  &  0.116 $\pm$ 0.154  &  2900  &   200  &  9.90  &  8.53  &  0.117 $\pm$ 0.169  \\
2800  &   300  &  2.87  &  2.55  &  0.235 $\pm$ 0.066  &  2900  &   300  &  3.25  &  2.80  &  0.227 $\pm$ 0.079  \\
2800  &   400  &  1.49  &  1.32  &  0.360 $\pm$ 0.036  &  2900  &   400  &  1.65  &  1.42  &  0.365 $\pm$ 0.045  \\
2800  &   500  &  0.992  &  0.879  &  0.482 $\pm$ 0.021  &  2900  &   500  &  1.07  &  0.919  &  0.471 $\pm$ 0.040  \\
2800  &   600  &  0.774  &  0.686  &  0.576 $\pm$ 0.016  &  2900  &   600  &  0.817  &  0.703  &  0.570 $\pm$ 0.017  \\
2800  &   700  &  0.654  &  0.580  &  0.645 $\pm$ 0.015  &  2900  &   700  &  0.688  &  0.593  &  0.637 $\pm$ 0.015  \\
\hline
\end{tabular}
\label{tab:muLimits2800to2900}
\end{table*}

\begin{table*}[!htbp]
\centering
\topcaption{The 95\% CL observed (Obs.) and expected (Exp.) exclusion limits (in fb) on the \WR production cross section times branching fraction for $\WR \to \mu \mu j j$ as a function of \WR and \Nmu mass (in \GeVns) for $3000 \leq \MWR \leq 3100$\GeV.  This signal acceptance (Acc.) is also included for each $(\MWR,M_{\Nmu})$ entry.}
\begin{tabular}{ccccc|ccccc}
\hline
\MWR & $M_{\Nmu}$ & Obs. & Exp. & Acc. & \MWR &  $M_{\Nmu}$ & Obs. & Exp. & Acc. \\
\hline
3000  &   100  &  72.8  &  63.2  &  0.024 $\pm$ 0.428  &  3100  &   100  &  75.6  &  65.9  &  0.029 $\pm$ 0.439  \\
3000  &   200  &  11.3  &  9.79  &  0.117 $\pm$ 0.185  &  3100  &   200  &  12.9  &  11.2  &  0.130 $\pm$ 0.197  \\
3000  &   300  &  3.59  &  3.12  &  0.234 $\pm$ 0.094  &  3100  &   300  &  4.15  &  3.62  &  0.240 $\pm$ 0.104  \\
3000  &   400  &  1.81  &  1.57  &  0.356 $\pm$ 0.052  &  3100  &   400  &  2.01  &  1.75  &  0.359 $\pm$ 0.060  \\
3000  &   500  &  1.16  &  1.01  &  0.469 $\pm$ 0.031  &  3100  &   500  &  1.28  &  1.12  &  0.461 $\pm$ 0.037  \\
3000  &   600  &  0.864  &  0.750  &  0.561 $\pm$ 0.019  &  3100  &   600  &  0.935  &  0.815  &  0.552 $\pm$ 0.024  \\
3000  &   700  &  0.714  &  0.620  &  0.627 $\pm$ 0.016  &  3100  &   700  &  0.763  &  0.665  &  0.623 $\pm$ 0.021  \\
\hline
\end{tabular}
\label{tab:muLimits3000to3100}
\end{table*}

\begin{table*}[!htbp]
\centering
\topcaption{The 95\% CL observed (Obs.) and expected (Exp.) exclusion limits (in fb) on the \WR production cross section times branching fraction for $\WR \to \mu \mu j j$ as a function of \WR and \Nmu mass (in \GeVns) for $ \MWR = 3200$\GeV.  This signal acceptance (Acc.) is also included for each $(\MWR,M_{\Nmu})$ entry.}
\begin{tabular}{ccccc}
\hline
\MWR & $M_{\Nmu}$ & Obs. & Exp. & Acc. \\
\hline
3200  &   100  &  75.8  &  66.0  &  0.031 $\pm$ 0.449  \\
3200  &   200  &  14.4  &  12.5  &  0.131 $\pm$ 0.206  \\
3200  &   300  &  4.58  &  3.99  &  0.254 $\pm$ 0.111  \\
3200  &   400  &  2.24  &  1.95  &  0.363 $\pm$ 0.070  \\
3200  &   500  &  1.39  &  1.21  &  0.458 $\pm$ 0.050  \\
3200  &   600  &  1.01  &  0.877  &  0.546 $\pm$ 0.029  \\
3200  &   700  &  0.808  &  0.704  &  0.614 $\pm$ 0.020  \\
3200  &   800  &  0.691  &  0.602  &  0.668 $\pm$ 0.016  \\
3200  &   900  &  0.614  &  0.535  &  0.703 $\pm$ 0.018  \\
3200  &  1000  &  0.567  &  0.494  &  0.734 $\pm$ 0.014  \\
3200  &  1100  &  0.531  &  0.463  &  0.762 $\pm$ 0.027  \\
3200  &  1200  &  0.503  &  0.438  &  0.776 $\pm$ 0.015  \\
3200  &  1300  &  0.485  &  0.422  &  0.790 $\pm$ 0.026  \\
3200  &  1400  &  0.468  &  0.408  &  0.799 $\pm$ 0.015  \\
3200  &  1500  &  0.455  &  0.396  &  0.808 $\pm$ 0.016  \\
3200  &  1600  &  0.445  &  0.388  &  0.819 $\pm$ 0.015  \\
3200  &  1700  &  0.444  &  0.387  &  0.820 $\pm$ 0.030  \\
3200  &  1800  &  0.441  &  0.384  &  0.827 $\pm$ 0.018  \\
3200  &  1900  &  0.440  &  0.383  &  0.829 $\pm$ 0.031  \\
3200  &  2000  &  0.438  &  0.382  &  0.832 $\pm$ 0.023  \\
3200  &  2100  &  0.442  &  0.385  &  0.825 $\pm$ 0.018  \\
3200  &  2200  &  0.438  &  0.382  &  0.831 $\pm$ 0.019  \\
3200  &  2300  &  0.441  &  0.385  &  0.826 $\pm$ 0.019  \\
3200  &  2400  &  0.443  &  0.386  &  0.822 $\pm$ 0.015  \\
3200  &  2500  &  0.446  &  0.389  &  0.817 $\pm$ 0.018  \\
3200  &  2600  &  0.449  &  0.391  &  0.811 $\pm$ 0.017  \\
3200  &  2700  &  0.451  &  0.393  &  0.808 $\pm$ 0.021  \\
3200  &  2800  &  0.459  &  0.399  &  0.795 $\pm$ 0.021  \\
3200  &  2900  &  0.471  &  0.410  &  0.774 $\pm$ 0.022  \\
3200  &  3000  &  0.485  &  0.423  &  0.751 $\pm$ 0.015  \\
3200  &  3100  &  0.527  &  0.459  &  0.691 $\pm$ 0.039  \\
\hline
\end{tabular}
\label{tab:muLimits3200to3200}
\end{table*}

\begin{table*}[!htbp]
\centering
\topcaption{The 95\% CL observed (Obs.) and expected (Exp.) exclusion limits (in fb) on the \WR production cross section times branching fraction for $\WR \to (\Pe \Pe + \mu \mu) j j$ as a function of \WR and \Nell mass (in \GeVns) for $1000 \leq \MWR \leq 1600$\GeV.  The signal acceptance (Acc.) is also included for each $(\MWR,M_{\Nell})$ entry.}
\begin{tabular}{ccccc|ccccc}
\hline
\MWR & $M_{\Nell}$ & Obs. & Exp. & Acc. & \MWR &  $M_{\Nell}$ & Obs. & Exp. & Acc. \\
\hline
1000  &   100  &  117  &  88.5  &  0.0695 $\pm$ 0.0650  &  1400  &   100  &  54.0  &  71.9  &  0.040 $\pm$ 0.109  \\
1000  &   200  &  26.3  &  19.9  &  0.291 $\pm$ 0.018  &  1400  &   200  &  8.26  &  11.0  &  0.232 $\pm$ 0.034  \\
1000  &   300  &  17.8  &  13.5  &  0.441 $\pm$ 0.010  &  1400  &   300  &  4.52  &  6.02  &  0.419 $\pm$ 0.016  \\
1000  &   400  &  15.0  &  11.3  &  0.544 $\pm$ 0.009  &  1400  &   400  &  3.49  &  4.64  &  0.541 $\pm$ 0.011  \\
1000  &   500  &  13.7  &  10.3  &  0.591 $\pm$ 0.009  &  1400  &   500  &  3.07  &  4.08  &  0.624 $\pm$ 0.009  \\
1000  &   600  &  13.1  &  9.90  &  0.616 $\pm$ 0.008  &  1400  &   600  &  2.89  &  3.84  &  0.682 $\pm$ 0.009  \\
1000  &   700  &  13.1  &  9.94  &  0.613 $\pm$ 0.008  &  1400  &   700  &  2.77  &  3.69  &  0.707 $\pm$ 0.009  \\
\hline
\end{tabular}
\label{tab:emuLimits1000to1600}
\end{table*}

\begin{table*}[!htbp]
\centering
\topcaption{The 95\% CL observed (Obs.) and expected (Exp.) exclusion limits (in fb) on the \WR production cross section times branching fraction for $\WR \to (\Pe \Pe + \mu \mu) j j$ as a function of \WR and \Nell mass (in \GeVns) for $1700 \leq \MWR \leq 2000$\GeV.  The signal acceptance (Acc.) is also included for each $(\MWR,M_{\Nell})$ entry.}
\begin{tabular}{ccccc|ccccc}
\hline
\MWR & $M_{\Nell}$ & Obs. & Exp. & Acc. & \MWR &  $M_{\Nell}$ & Obs. & Exp. & Acc. \\
\hline
1700  &   100  &  60.3  &  75.0  &  0.029 $\pm$ 0.140  &  1900  &   100  &  117  &  69.5  &  0.022 $\pm$ 0.159  \\
1700  &   200  &  7.31  &  9.09  &  0.187 $\pm$ 0.049  &  1900  &   200  &  13.0  &  7.71  &  0.154 $\pm$ 0.061  \\
1700  &   300  &  3.13  &  3.90  &  0.373 $\pm$ 0.023  &  1900  &   300  &  5.37  &  3.17  &  0.338 $\pm$ 0.031  \\
1700  &   400  &  2.21  &  2.75  &  0.516 $\pm$ 0.014  &  1900  &   400  &  3.63  &  2.14  &  0.492 $\pm$ 0.018  \\
1700  &   500  &  1.88  &  2.34  &  0.609 $\pm$ 0.010  &  1900  &   500  &  3.01  &  1.78  &  0.596 $\pm$ 0.015  \\
1700  &   600  &  1.76  &  2.19  &  0.670 $\pm$ 0.009  &  1900  &   600  &  2.71  &  1.60  &  0.667 $\pm$ 0.010  \\
1700  &   700  &  1.72  &  2.14  &  0.714 $\pm$ 0.009  &  1900  &   700  &  2.58  &  1.52  &  0.709 $\pm$ 0.009  \\
\hline
\end{tabular}
\label{tab:emuLimits1700to2000}
\end{table*}

\begin{table*}[!htbp]
\centering
\topcaption{The 95\% CL observed (Obs.) and expected (Exp.) exclusion limits (in fb) on the \WR production cross section times branching fraction for $\WR \to (\Pe \Pe + \mu \mu) j j$ as a function of \WR and \Nell mass (in \GeVns) for $2100 \leq \MWR \leq 2300$\GeV.  The signal acceptance (Acc.) is also included for each $(\MWR,M_{\Nell})$ entry.}
\begin{tabular}{ccccc|ccccc}
\hline
\MWR & $M_{\Nell}$ & Obs. & Exp. & Acc. & \MWR &  $M_{\Nell}$ & Obs. & Exp. & Acc. \\
\hline
2100  &   100  &  172  &  79.3  &  0.022 $\pm$ 0.178  &  2300  &   100  &  147  &  77.7  &  0.020 $\pm$ 0.197  \\
2100  &   200  &  18.5  &  8.55  &  0.135 $\pm$ 0.070  &  2300  &   200  &  14.8  &  7.81  &  0.127 $\pm$ 0.082  \\
2100  &   300  &  7.14  &  3.30  &  0.305 $\pm$ 0.034  &  2300  &   300  &  5.57  &  2.94  &  0.286 $\pm$ 0.041  \\
2100  &   400  &  4.41  &  2.04  &  0.458 $\pm$ 0.019  &  2300  &   400  &  3.33  &  1.76  &  0.425 $\pm$ 0.023  \\
2100  &   500  &  3.39  &  1.57  &  0.576 $\pm$ 0.016  &  2300  &   500  &  2.56  &  1.35  &  0.543 $\pm$ 0.015  \\
2100  &   600  &  2.90  &  1.34  &  0.648 $\pm$ 0.010  &  2300  &   600  &  2.23  &  1.18  &  0.641 $\pm$ 0.011  \\
2100  &   700  &  2.65  &  1.22  &  0.703 $\pm$ 0.009  &  2300  &   700  &  2.03  &  1.07  &  0.683 $\pm$ 0.010  \\
\hline
\end{tabular}
\label{tab:emuLimits2100to2300}
\end{table*}

\begin{table*}[!htbp]
\centering
\topcaption{The 95\% CL observed (Obs.) and expected (Exp.) exclusion limits (in fb) on the \WR production cross section times branching fraction for $\WR \to (\Pe \Pe + \mu \mu) j j$ as a function of \WR and \Nell mass (in \GeVns) for $2400 \leq \MWR \leq 2500$\GeV.  The signal acceptance (Acc.) is also included for each $(\MWR, M_{\Nell})$ entry.}
\begin{tabular}{ccccc|ccccc}
\hline
\MWR & $M_{\Nell}$ & Obs. & Exp. & Acc. & \MWR &  $M_{\Nell}$ & Obs. & Exp. & Acc. \\
\hline
2400  &   100  &  151  &  88.1  &  0.022 $\pm$ 0.207  &  2500  &   100  &  153  &  98.3  &  0.024 $\pm$ 0.217  \\
2400  &   200  &  14.0  &  8.18  &  0.121 $\pm$ 0.088  &  2500  &   200  &  13.8  &  8.85  &  0.119 $\pm$ 0.095  \\
2400  &   300  &  5.00  &  2.93  &  0.268 $\pm$ 0.044  &  2500  &   300  &  4.71  &  3.02  &  0.254 $\pm$ 0.049  \\
2400  &   400  &  2.90  &  1.69  &  0.418 $\pm$ 0.032  &  2500  &   400  &  2.67  &  1.71  &  0.401 $\pm$ 0.032  \\
2400  &   500  &  2.15  &  1.26  &  0.537 $\pm$ 0.016  &  2500  &   500  &  1.88  &  1.21  &  0.527 $\pm$ 0.018  \\
2400  &   600  &  1.79  &  1.05  &  0.624 $\pm$ 0.012  &  2500  &   600  &  1.56  &  0.999  &  0.612 $\pm$ 0.013  \\
2400  &   700  &  1.63  &  0.956  &  0.683 $\pm$ 0.010  &  2500  &   700  &  1.38  &  0.884  &  0.674 $\pm$ 0.011  \\
\hline
\end{tabular}
\label{tab:emuLimits2400to2500}
\end{table*}

\begin{table*}[!htbp]
\centering
\topcaption{The 95\% CL observed (Obs.) and expected (Exp.) exclusion limits (in fb) on the \WR production cross section times branching fraction for $\WR \to (\Pe \Pe + \mu \mu) j j$ as a function of \WR and \Nell mass (in \GeVns) for $2600 \leq \MWR \leq 2700$\GeV.  The signal acceptance (Acc.) is also included for each $(\MWR,M_{\Nell})$ entry.}
\begin{tabular}{ccccc|ccccc}
\hline
\MWR & $M_{\Nell}$ & Obs. & Exp. & Acc. & \MWR &  $M_{\Nell}$ & Obs. & Exp. & Acc. \\
\hline
2600  &   100  &  136  &  98.1  &  0.021 $\pm$ 0.225  &  2700  &   100  &  134  &  94.6  &  0.023 $\pm$ 0.236  \\
2600  &   200  &  13.3  &  9.58  &  0.115 $\pm$ 0.101  &  2700  &   200  &  14.1  &  9.94  &  0.120 $\pm$ 0.109  \\
2600  &   300  &  4.42  &  3.19  &  0.246 $\pm$ 0.053  &  2700  &   300  &  4.70  &  3.31  &  0.247 $\pm$ 0.058  \\
2600  &   400  &  2.38  &  1.72  &  0.392 $\pm$ 0.031  &  2700  &   400  &  2.48  &  1.75  &  0.381 $\pm$ 0.034  \\
2600  &   500  &  1.66  &  1.20  &  0.509 $\pm$ 0.019  &  2700  &   500  &  1.70  &  1.20  &  0.500 $\pm$ 0.021  \\
2600  &   600  &  1.34  &  0.966  &  0.605 $\pm$ 0.014  &  2700  &   600  &  1.34  &  0.940  &  0.601 $\pm$ 0.015  \\
2600  &   700  &  1.17  &  0.842  &  0.672 $\pm$ 0.011  &  2700  &   700  &  1.18  &  0.831  &  0.663 $\pm$ 0.012  \\
\hline
\end{tabular}
\label{tab:emuLimits2600to2700}
\end{table*}

\begin{table*}[!htbp]
\centering
\topcaption{The 95\% CL observed (Obs.) and expected (Exp.) exclusion limits (in fb) on the \WR production cross section times branching fraction for $\WR \to (\Pe \Pe + \mu \mu) j j$ as a function of \WR and \Nell mass (in \GeVns) for $2800 \leq \MWR \leq 2900$\GeV.  The signal acceptance (Acc.) is also included for each $(\MWR,M_{\Nell})$ entry.}
\begin{tabular}{ccccc|ccccc}
\hline
\MWR & $M_{\Nell}$ & Obs. & Exp. & Acc. & \MWR &  $M_{\Nell}$ & Obs. & Exp. & Acc. \\
\hline
2800  &   100  &  143  &  101  &  0.025 $\pm$ 0.248  &  2900  &   100  &  140  &  100  &  0.027 $\pm$ 0.255  \\
2800  &   200  &  16.5  &  11.6  &  0.123 $\pm$ 0.117  &  2900  &   200  &  18.1  &  12.9  &  0.123 $\pm$ 0.126  \\
2800  &   300  &  5.46  &  3.86  &  0.245 $\pm$ 0.063  &  2900  &   300  &  5.90  &  4.21  &  0.236 $\pm$ 0.070  \\
2800  &   400  &  2.81  &  1.99  &  0.372 $\pm$ 0.037  &  2900  &   400  &  2.97  &  2.12  &  0.377 $\pm$ 0.042  \\
2800  &   500  &  1.86  &  1.32  &  0.494 $\pm$ 0.023  &  2900  &   500  &  1.91  &  1.37  &  0.485 $\pm$ 0.030  \\
2800  &   600  &  1.45  &  1.02  &  0.589 $\pm$ 0.016  &  2900  &   600  &  1.46  &  1.04  &  0.582 $\pm$ 0.017  \\
2800  &   700  &  1.22  &  0.861  &  0.655 $\pm$ 0.013  &  2900  &   700  &  1.22  &  0.873  &  0.647 $\pm$ 0.014  \\
\hline
\end{tabular}
\label{tab:emuLimits2800to2900}
\end{table*}

\begin{table*}[!htbp]
\centering
\topcaption{The 95\% CL observed (Obs.) and expected (Exp.) exclusion limits (in fb) on the \WR production cross section times branching fraction for $\WR \to (\Pe \Pe + \mu \mu) j j$ as a function of \WR and \Nell mass (in \GeVns) for $3000 \leq \MWR \leq 3100$\GeV.  The signal acceptance (Acc.) is also included for each $(\MWR,M_{\Nell})$ entry.}
\begin{tabular}{ccccc|ccccc}
\hline
\MWR & $M_{\Nell}$ & Obs. & Exp. & Acc. & \MWR &  $M_{\Nell}$ & Obs. & Exp. & Acc. \\
\hline
3000  &   100  &  137  &  99.7  &  0.027 $\pm$ 0.264  &  3100  &   100  &  141  &  103  &  0.032 $\pm$ 0.269  \\
3000  &   200  &  20.3  &  14.7  &  0.124 $\pm$ 0.135  &  3100  &   200  &  23.0  &  16.7  &  0.137 $\pm$ 0.142  \\
3000  &   300  &  6.40  &  4.66  &  0.245 $\pm$ 0.078  &  3100  &   300  &  7.33  &  5.32  &  0.250 $\pm$ 0.086  \\
3000  &   400  &  3.21  &  2.33  &  0.369 $\pm$ 0.048  &  3100  &   400  &  3.53  &  2.56  &  0.373 $\pm$ 0.054  \\
3000  &   500  &  2.04  &  1.49  &  0.482 $\pm$ 0.030  &  3100  &   500  &  2.24  &  1.63  &  0.475 $\pm$ 0.034  \\
3000  &   600  &  1.52  &  1.10  &  0.573 $\pm$ 0.019  &  3100  &   600  &  1.63  &  1.18  &  0.565 $\pm$ 0.022  \\
3000  &   700  &  1.25  &  0.908  &  0.637 $\pm$ 0.014  &  3100  &   700  &  1.32  &  0.959  &  0.634 $\pm$ 0.018  \\
\hline
\end{tabular}
\label{tab:emuLimits3000to3100}
\end{table*}

\begin{table*}[!htbp]
\centering
\topcaption{The 95\% CL observed (Obs.) and expected (Exp.) exclusion limits (in fb) on the \WR production cross section times branching fraction for $\WR \to (\Pe \Pe + \mu \mu) j j$ as a function of \WR and \Nell mass (in \GeVns) for $ \MWR = 3200$\GeV.  The signal acceptance (Acc.) is also included for each $(\MWR,M_{\Nell})$ entry.}
\begin{tabular}{ccccc}
\hline
\MWR & $M_{\Nell}$ & Obs. & Exp. & Acc. \\
\hline
3200  &   100  &  144  &  104  &  0.035 $\pm$ 0.274  \\
3200  &   200  &  25.9  &  18.7  &  0.139 $\pm$ 0.147  \\
3200  &   300  &  8.16  &  5.90  &  0.265 $\pm$ 0.091  \\
3200  &   400  &  3.98  &  2.88  &  0.376 $\pm$ 0.059  \\
3200  &   500  &  2.46  &  1.78  &  0.471 $\pm$ 0.041  \\
3200  &   600  &  1.77  &  1.28  &  0.560 $\pm$ 0.025  \\
3200  &   700  &  1.41  &  1.02  &  0.626 $\pm$ 0.019  \\
3200  &   800  &  1.20  &  0.871  &  0.679 $\pm$ 0.014  \\
3200  &   900  &  1.07  &  0.772  &  0.712 $\pm$ 0.016  \\
3200  &  1000  &  0.983  &  0.711  &  0.741 $\pm$ 0.012  \\
3200  &  1100  &  0.920  &  0.665  &  0.768 $\pm$ 0.016  \\
3200  &  1200  &  0.870  &  0.629  &  0.781 $\pm$ 0.010  \\
3200  &  1300  &  0.838  &  0.606  &  0.795 $\pm$ 0.016  \\
3200  &  1400  &  0.809  &  0.585  &  0.804 $\pm$ 0.013  \\
3200  &  1500  &  0.785  &  0.568  &  0.813 $\pm$ 0.011  \\
3200  &  1600  &  0.769  &  0.556  &  0.823 $\pm$ 0.010  \\
3200  &  1700  &  0.767  &  0.555  &  0.824 $\pm$ 0.016  \\
3200  &  1800  &  0.760  &  0.550  &  0.831 $\pm$ 0.013  \\
3200  &  1900  &  0.758  &  0.548  &  0.832 $\pm$ 0.016  \\
3200  &  2000  &  0.756  &  0.547  &  0.834 $\pm$ 0.020  \\
3200  &  2100  &  0.762  &  0.551  &  0.828 $\pm$ 0.010  \\
3200  &  2200  &  0.756  &  0.547  &  0.835 $\pm$ 0.013  \\
3200  &  2300  &  0.762  &  0.551  &  0.829 $\pm$ 0.011  \\
3200  &  2400  &  0.765  &  0.553  &  0.825 $\pm$ 0.009  \\
3200  &  2500  &  0.770  &  0.557  &  0.819 $\pm$ 0.010  \\
3200  &  2600  &  0.775  &  0.560  &  0.814 $\pm$ 0.010  \\
3200  &  2700  &  0.778  &  0.563  &  0.811 $\pm$ 0.013  \\
3200  &  2800  &  0.791  &  0.572  &  0.797 $\pm$ 0.012  \\
3200  &  2900  &  0.813  &  0.588  &  0.776 $\pm$ 0.013  \\
3200  &  3000  &  0.837  &  0.605  &  0.753 $\pm$ 0.010  \\
3200  &  3100  &  0.909  &  0.658  &  0.693 $\pm$ 0.024  \\
\hline
\end{tabular}
\label{tab:emuLimits3200to3200}
\end{table*}

}
\cleardoublepage \section{The CMS Collaboration \label{app:collab}}\begin{sloppypar}\hyphenpenalty=5000\widowpenalty=500\clubpenalty=5000\textbf{Yerevan Physics Institute,  Yerevan,  Armenia}\\*[0pt]
V.~Khachatryan, A.M.~Sirunyan, A.~Tumasyan
\vskip\cmsinstskip
\textbf{Institut f\"{u}r Hochenergiephysik der OeAW,  Wien,  Austria}\\*[0pt]
W.~Adam, T.~Bergauer, M.~Dragicevic, J.~Er\"{o}, C.~Fabjan\cmsAuthorMark{1}, M.~Friedl, R.~Fr\"{u}hwirth\cmsAuthorMark{1}, V.M.~Ghete, C.~Hartl, N.~H\"{o}rmann, J.~Hrubec, M.~Jeitler\cmsAuthorMark{1}, W.~Kiesenhofer, V.~Kn\"{u}nz, M.~Krammer\cmsAuthorMark{1}, I.~Kr\"{a}tschmer, D.~Liko, I.~Mikulec, D.~Rabady\cmsAuthorMark{2}, B.~Rahbaran, H.~Rohringer, R.~Sch\"{o}fbeck, J.~Strauss, A.~Taurok, W.~Treberer-Treberspurg, W.~Waltenberger, C.-E.~Wulz\cmsAuthorMark{1}
% ...
\vskip\cmsinstskip
% ...
1:~~Also at Vienna University of Technology, Vienna, Austria\\
2:~~Also at CERN, European Organization for Nuclear Research, Geneva, Switzerland\\
% ...

\end{sloppypar}
\end{document}